\documentclass[pra, two column, superscriptaddress, showpacs,nofootinbib,longbibliography, title page]{revtex4-2}

\usepackage{physics}
\usepackage{enumerate}
\usepackage{etex}
\usepackage{amsmath,amssymb,amsthm,amstext, mathrsfs}
\usepackage[pdftex]{graphicx}
\usepackage{verbatim}
\usepackage{bbold}
\usepackage{mathtools}
\usepackage{pgfplots}
\pgfplotsset{width=8 cm,compat=1.8}
\usepackage{times,txfonts}
\usepackage{braket}
\usepackage{color}
\usepackage{array}
\usepackage{blkarray}
\usepackage{mathtools}
\usepackage{latexsym}
\usepackage{tabularx, booktabs}
\usepackage{graphics,epstopdf}
\usepackage{graphicx}
\usepackage{float}
\usepackage{amsfonts}
\usepackage{titlesec}
\usepackage{subfigure}
\usepackage{hyperref}
\usepackage{mathrsfs}
\hypersetup{
	colorlinks=true,
	linkcolor=red,
	filecolor=blue,      
	urlcolor=blue,
	citecolor=purple,
}

\DeclareMathOperator*{\Motimes}{\text{\raisebox{0.25ex}{\scalebox{0.65}{$\bigotimes$}}}}

\newtheorem{thm}{Theorem}

\begin{document}

\title{Probing the robustness of  various self-testing protocols for multipartite entangled states}

\author{Priyaranjan K. Jha}
\email{priyaranjankumarjha118@gmail.com}
\affiliation{Department of Physics, Indian Institute of Technology Hyderabad, Telangana-502284, India }

\author{Ritesh K. Singh}
\email{riteshcsm1@gmail.com}
\affiliation{Department of Physics, Indian Institute of Technology Hyderabad, Telangana-502284, India }

\author{A. K. Pan}
	\email{akp@phy.iith.ac.in}
	 \affiliation{Department of Physics, Indian Institute of Technology Hyderabad, Telangana-502284, India }


\begin{abstract}
 Device-independent certification of multipartite entangled states plays a central role 
 in a wide range of practical applications, including quantum networks, conference key agreement, and verifiable distributed quantum computation. A particularly important class of multipartite entangled states is the class of Greenberger-Horne-Zeilinger (GHZ) states. Many Bell operators have been proposed to self-test GHZ states. However, in practical scenarios, due to imperfections and the finite collection of statistics, the observed statistics do not satisfy the ideal self-testing relations. Hence, it becomes essential to investigate and compare the robustness of the different self-testing protocols. In this work, we investigate the robustness of self-testing schemes constructed from Bell operators due to Svetlichny and Mermin--Ardehali--Belinskii--Klyshko (MABK), using the analytic operator-inequality framework developed by Kaniewski [\href{https://doi.org/10.1103/PhysRevLett.117.070402}{Phys. Rev. Lett. 117, 070402 (2016)}]. We derive lower bounds on the extractable fidelity as a function of the observed value of these Bell operators. Although these protocols self-test the same underlying state, they exhibit markedly different levels of robustness. By comparing the resulting fidelity bounds, we demonstrate that the self-testing scheme based on the Svetlichny's Bell operator is the more robust among the two. Our results thus identify the Svetlichny operator based self-testing protocol as the most favorable candidate for device-independent certification of GHZ states in realistic, noisy experimental scenarios.
\end{abstract}

\maketitle

\section{Introduction}

Self-testing is a method to certify quantum systems based solely on  input-output statistics, allowing a classical verifier to certify quantum systems \cite{Supic2020} without any knowledge of the internal workings of the measurement devices and sources involved in an experiment. A common method of such certification involves a Bell scenario in which the observed statistics must maximally violate a Bell inequality, establishing that the observed correlations are extremal \cite{Mayers2004,McKague2012}. This type of certification is said to be device-independent since no assumptions about the devices involved in the experiment are made. In an ideal scenario, to self-test the states and measurements, in addition to the maximal violation of a Bell inequality, it is necessary to demonstrate the existence of a swap isometry from a physical experiment to a reference experiment. 

However, in experiments, the optimal quantum  violation of Bell inequalities cannot be observed due to the presence of noise, the collection of a finite number of statistics, etc. Hence, robust protocols for self-testing are required to analyze the extent to which the self-testing is valid. Early attempts at robust self-testing of maximally entangled state were in \cite{McKague2012} and these methodologies were extended to construct robust self-testing protocols for Pauli observables \cite{Bowels_Acin_2018}, multipartite entangled states \cite{McKague2016a,Singh2025a}, graph states \cite{McKague2016a}, etc. There were also some further results on robustly self-testing many copies of maximally  entangled states \cite{McKague2016,Wu2016,McKague2017,Coladangelo2017b}. 

These types of robustness analysis give non-linear bounds ~\cite{McKague2012,wu2014,Bamps2015,supic2016,McKague2018} of the difference between the ideal state and the extracted state via a swap isometry. These bounds are not tight, and if such methods are extended to multipartite scenarios \cite{Singh2025a}, they lead to a very limited robustness analysis due to the the repeated use of norm inequalities. Such robustness results are stable for only a very small deviations from the maximal violation of the considered Bell inequalities and are not experimentally relevant \cite{Bardyn2009,Kanewski2017}. A result that is usable in real experiments was obtained by using the swap method \cite{Yang2014, Bancal2015}, through the semi-definite optimization technique to bound the singlet fidelity with the extracted state from a extraction channel. In \cite{Yang2014}, it was shown that a violation of the Clauser–Horne–Shimony–Holt (CHSH) inequality greater than 2.57 certifies a state that has more than half fidelity with the singlet. In such swap methods, black boxes are replaced by ideal systems of known finite dimensions, and the error is considered to originate from the imperfect swap operators extracting a state different from the ideal state.

In contrast to the previously discussed works, Kaniewski \cite{Kaniewski2016} introduced an analytical method to robustly self-test the singlet state where the fidelity is not lower bounded by a  non-linear function of observed CHSH violation. He demonstrated that any CHSH \cite{Clauser1969} violation exceeding $2.11$ guarantees a singlet extraction fidelity larger than half. The same analytical framework was subsequently employed to robustly self-test the tripartite GHZ state via the Mermin inequality\cite{Mermin1990}. A  similar method was adapted \cite{Coopmans2019} for tilted Bell inequalities, numerically demonstrating robust self-testing for any arbitrary two-qubit states. In \cite{Coopmans2019}, it was further noted that robust self-testing of the singlet via CHSH is subject to a threshold, one can only make a self-testing claim for violations greater than $2.11$. This stands in contrast to the Mermin inequality, which does not exhibit such a non-trivial threshold. However, Mermin inequality does not exhibit genuine multipartite nonlocality.  In \cite{Baccari2020}, in a network scenario,  a protocol for robust self-testing of $n-$partite GHZ and $n-$partite ring states for $n\leq7$ was demonstrated using  numerical analysis.

In this work, building on the analytical framework introduced by Kaniewski \cite{Kaniewski2016}, we robustly self-test the GHZ state based on the Svetlichny and MABK inequalities. Unlike the original approach \cite{Kaniewski2016, Coopmans2019}, which requires demonstrating the positivity of intricate nonlinear parameter-dependent functions to verify an operator inequality, our method establishes this operator inequality by leveraging symmetry properties and designing a suitable unitary transformation. This construction streamlines the analysis and yields fully analytical robustness bounds as functions of the observed violations of the Svetlichny or MABK inequalities. We provide explicit analytical results for $3, 4 \ \text{and }5$-partite GHZ states and argue that our method can be generalized to an arbitrary $n$ number of parties.  We find that the self-testing protocol relying on the Svetlichny inequality offers the strongest robustness. We demonstrate that for any nonzero violation of the Svetlichny inequality, it is possible to extract a state with fidelity greater than half with respect to the ideal GHZ state.

The paper is organized as follows. In Sec. \ref{Sec:Preliminaries}, we introduce the Svetlichny  and MABK inequalities, and then the notion of self-testing. In Sec. \ref{Sec:STOPI} and Sec. \ref{Sec III} we build the tools required for our work and then in Sec. \ref{Sec IV}, we derive analytical robustness bounds for the $3$-partite, $4$-partite, and $5$-partite Svetlichny inequalities. In Sec. \ref{Sec V}, we obtain robustness estimates for a self-testing protocol based on the MABK inequalities. We then compare our findings with existing results in the literature and demonstrate that the robustness bounds derived for the $4$- and $5$-partite Svetlichny and MABK inequalities are tight.

\section{Preliminaries}{\label{Sec:Preliminaries}}
We encapsulate the correlation in a multipartite Bell scenario and introduce the Svetlichny and MABK inequalities. We then briefly discuss the idea of self-testing.  

\subsection{Multipartite Bell Scenario}

We consider a multipartite Bell scenario comprising $n$ spatially separated parties, denoted by $A_j$  ($j = 1,\ldots,n$), sharing a quantum state $\rho_n$ prepared by a common source. Each party has possible inputs, labeled by $x_j $, with outcomes denoted by $a_j$. The correlations observed by the parties in such a Bell experiment are encoded in a vector commonly referred to as a \emph{behavior}, defined as
\[
\vec{\mathbf{p}} = \left\{ p(\vec a \mid \vec{x}) \; \big| \;
\vec{a} = (a_1,a_2,\ldots,a_n),\;
\vec{x} = (x_1,x_2,\ldots,x_n) \right\}.
\]

Here, $p(\vec a \mid \vec x)$ denotes the joint probability of obtaining the set of outcomes $\vec a$ given that the parties performed the set of measurements $\vec x$.

In quantum theory, the joint probability distribution is given by
\begin{equation}
\vec{\mathbf{p}}_Q(\vec{a} \mid \vec{x})
=
\operatorname{Tr}
\left[
\rho_n
\left(
E^{a_1}_{x_1}
\otimes
E^{a_2}_{x_2}
\otimes
\cdots
\otimes
E^{a_n}_{x_n}
\right)
\right],
\label{eq:Quantum Expectation}
\end{equation}
where $E^{a_j}_{x_j}\in \mathcal{L}(\mathcal{H}_{A_i})$ are the POVM elements corresponding to outcome $a_j$ when the observable $A_j^{x_j}$ is measured.

A set of correlations admits a \emph{local realistic} (or local hidden variable) description if it satisfies
\begin{equation}
\vec{\mathbf{p}}(\vec{a} \mid \vec{x})
=
\int d\lambda \, \mu(\lambda)
\prod_{j=1}^{n}
\xi(a_j\mid x_j, \lambda)
\label{eq:Local Model}
\end{equation}
Here, $\lambda$ denotes the hidden variable that determines the outcomes of the measurements of each party, and $\mu(\lambda)$ is a normalized probability distribution over $\lambda$. The function $\xi(a_j| x_j, \lambda)$ is the local response function which may be assumed to be deterministic. Any correlation not admitting a model of the form in Eq.~(\ref{eq:Local Model}) is called nonlocal correlation.

 In this work, we consider a multipartite Bell scenarios involving $n$ spatially separated parties, where each party performs two dichotomic measurements $x_j \in \{0,1\}$ with possible outcomes $a_j \in \{-1,+1\}$. We mainly focus on the Svetlichny and the MABK operators in this scenario. The $n$-party Svetlichny operator is given by \cite{Singh2025a}
\begin{equation}
\mathcal{S}_n
=
\sum_{\mu=1}^{2^{n-1}}
\left[
\nu_\mu^{\,m}
\left(
A_0^{1}
+
(-1)^{n+1} A_1^{1}
\right)
\bigotimes_{j=2}^{n}
A^{x_{\mu,j}}_{j}
\right]
\label{eq:SvOperator}
\end{equation}
 To fix $\nu_\mu^m$ and $x_{\mu,j}$ we define a $(2^{n-1}\times n)$ dimensional matrix $M$, whose each row represents one of possible $2^{n-1}$ number of $n$-bit strings. An entry of this matrix is represented by $x_{\mu,j}$, where $\mu$ represents the row index and $j$ is the column index. The entries of a particular row $\mu$ are determined as follows. For each increment in the index $\mu$, we add binary number $1$ to the binary number $x_\mu$ so that $x_{\mu+1} = x_\mu +_b 1$, where $+_b$ is addition operation on binary numbers. Starting with, $\mu = 1$, (\textit{i.e.}, $1$st row in the defined matrix $M$), $x^j_1 = 0 \ \forall j\in[n-1]$. For $\mu=2$, $x^k_2 = 0 \ \forall k\in [n-2]$ and $x^{n-1}_2 = 1$. This procedure of adding binary number $1$ to the previous entry is repeated till $\mu = 2^{n-1}$. Any row $\mu$ contains $m$ number of $1$'s and for that particular $\mu$, coefficient $v^m_{\mu}$ is defined as $v^m_{\mu} = (-1)^{m(m+1)/2} \in \{-1,1\}$. For the Svetlichny operator $\mathcal{S}_n$, the maximum local value $(\mathcal{S}_n)_L = 2^{n-1}$ and the optimal quantum value $(\mathcal{S}_n)_Q = 2^{n-1}\sqrt{2}$ \cite{Singh2025a,Svetlichny1987,Seevinck2002}.

The $n$-party MABK operator can be obtained using the compact expression \cite{Panwar2023}
\begin{equation}
\begin{aligned}
\mathcal{M}_n
=
&\frac{1}{2}\Biggl[\left( \frac{1-i}{\sqrt{2}} \right)^{(n-1)\oplus 2}
\bigotimes_{j=1}^{n}
\left( A_j^{0} + i A_j^{1} \right) \\
&+
\left( \frac{1+i}{\sqrt{2}} \right)^{(n-1)\oplus 2}
\bigotimes_{j=1}^{n}
\left( A_j^{0} - i A_j^{1}\right)\Biggr]
\end{aligned}
\label{eq:MABK Operator}
\end{equation}
where $A_j^{0}$ and $A_j^{1}$ denote the two dichotomic observables measured by the $j$th party, for which the local bound is $(\mathcal{M}_n)_L = 2^{\frac{n-1}{2}}$ and the optimal quantum value is $(\mathcal{M}_n)_Q = 2^{n-1}$.
\subsection{Self-testing}
An observed extremal correlation $\vec{\mathbf{p}}(\vec{a} \mid \vec{x})$ produced by a \textit{physical experiment} constituting physical state $\rho_n$ and physical measurements $A_j^{x_j}$, self-tests a known multipartite entangled state $\Psi'_n = \ket{\psi'_n}\bra{\psi'_n}$ in a \textit{reference experiment} also producing $\vec{\mathbf{p}}(\vec{a} \mid \vec{x})$, if there exists a local isometry such that, when applied to a purification $\ket{\psi}_{nP}$ of $\rho_n$, it extracts an exact copy of $\ket{\psi'}_{n}$~\cite{Supic2020}, i.e.,
\begin{equation}
    \left( \bigotimes_{j=1}^{n}{\Phi_{A_j}} \right)
    \otimes \mathbb{I}_P
    \left[\, \ket{\psi}_{nP} \,\right]
    =
    \ket{\psi'}_{n} \otimes \ket{\xi}_{\text{junk}}
    \label{eq:vector self test}
\end{equation}
In \cite{Coopmans2019} it is shown that this formulation is equivalent to saying that $\rho_{n}$ contains $\ket{\psi'}_{n}$ if there exist local quantum channels $\Lambda_{A_j} : \mathcal{L}(\mathcal{H}_{A_j})\rightarrow\mathcal{L}(\mathcal{H}_{{A_j}'})$, such that we can extract a perfect copy of $\Psi'_{n}$ from $\rho_{n}$, i.e.,
\begin{equation}
    \mathbf{\Lambda}\rho_{n} = \Psi_{n}' \qquad \text{where} \qquad \mathbf{\Lambda}=\bigotimes_{j=1}^{n}\Lambda_{A_j}
\label{eq:matrix self test}
\end{equation}
Operationally, self-testing implies that if an extremal correlation is generated by two different states $\rho_n$ and $\Psi^\prime_n$, (by the use of appropriate measurments) then, $\rho_n$ and $\Psi_n^\prime$ are related via a valid isometric channel.

\section{Robust Self-Testing from Operator Inequality}\label{Sec:STOPI}
The self-testing statement defined in Eqs. (\ref{eq:vector self test}) and (\ref{eq:matrix self test}) above cannot be used directly in real experiments, as the experiments performed in practice always have some level of noise, and we can never obtain perfect correlation since we only have a finite sample on hand. So, we need a self-testing statement that can certify observed statistics as being ``close" to the ideal statistics. This approach to self-test practical statistics is known as robust self-testing \cite{McKague2012}. A usual method for quantifying robustness is the trace-distance method \cite{McKague2012,McKague2016,Bowles2018}, which quantifies the trace distance of the extracted state from the ideal state. Such quantification yields robustness bounds as nonlinear functions of observed violation. However, these bounds are not tight, which makes these methods inapplicable to real experimental scenarios, as they require extremely small deviations from the ideal statistics.

In this paper, we use the self-testing from operator inequality (STOPI) method, introduced in \cite{Kaniewski2016} and expanded in \cite{Coopmans2019}. Such an approach yields stringent bounds on the fidelity between the extracted state and the ideal state, expressed as a linear function of the observed violation. Note that the STOPI method was originally introduced for bipartite Bell scenario. We extend this analytical approach to the  multipartite scenario.

\subsection{Obtaining the Robust self-testing statement}To establish the robust self-testing of multipartite state $\Psi'_n$ consider a Bell functional $\mathscr{B}$ with the optimal local  value $\beta_L$ and optimal quantum value $\beta_Q > \beta_L$. If the value of the Bell functional $\mathscr{B}$ is observed to be optimal quantum bound $\beta_Q$, the state $\Psi'_n$ can be perfectly self-tested. But perfect correlations are unattainable in realistic scenarios, hence it is necessary to generalize the notion of extraction introduced in Eq.~(\ref{eq:matrix self test}), which quantifies how closely the observed statistics approximate the ideal case. To this end, we define the extractability map $\Xi$ of the target state $\Psi'_{n}$ from  any arbitrary non ideal state $\rho_n$ as~\cite{Kaniewski2016, Bardyn2009}
\begin{equation}
    \Xi(\rho_{n}\rightarrow\Psi'_{n}) = \max_{\mathbf{\Lambda}} F(\mathbf{\Lambda}(\rho_{n}), \Psi'_{n}).
\label{eq:extractibility}
\end{equation}
where $F(\rho,\sigma)= \norm{\sqrt{\rho}\sqrt{\sigma}}_1$ is the fidelity. The maximization of fidelity $F$ is performed over all quantum channels mapping $\mathcal{L}(A_j)$ to $\mathcal{L}(A_j')$. To obtain the optimal robust self-testing statement, one has to minimize $\Xi$ over set of all states $\rho_n$ such that $\Tr{\mathscr{B}\rho_n}=\beta_O$ where $\beta_O$ is the observed violation of the Bell functional $\mathscr{B}$ to obtain the tradeoff between extractability and Bell violation. Such a tradeoff function $\mathcal{Q}$ is defined as
\begin{equation}
    \mathcal{Q}\qty[{\Psi'_n,\mathscr{B},\beta_O}] = \inf_{\rho_n}\Xi(\rho_n\rightarrow\Psi'_n)
\label{eq:tradeoff}
\end{equation}
 where $\beta_O >\beta_L$. This minimization problem is  quite challenging and the general method to finding the self-testing statement consists of fixing the extraction isometry and minimizing the fidelity only. In this work we consider the dephasing channel \cite{Kaniewski2016} and compare the robustness for various Bell type functionals.
 Now, noticing that any $\beta_O \in [\beta_L,\beta_Q]$ can be written as,
 \begin{equation}
     \beta_O = p\beta_Q + (1-p)\beta_L
 \label{eq:liner beta}
 \end{equation}
 for some $p \in [0,1]$, then $\beta_O$ can be obtained from the state
 \begin{equation}
     \rho_n =\Psi'_n \Motimes_{i=1}^n \ket{0}\bra{0}_{A_i} + \sigma_n\Motimes_{i=1}^n \ket{1}\bra{1}_{A_i}
 \label{eq:State Equation}
 \end{equation}
where $\sigma_n$ is an arbitrary separable state. Then the extractability of $\Psi'_n$ from $\rho_n$ is bounded by
\begin{equation}
    \Xi(\rho_n \rightarrow \Psi'_n) \leq p + (1-p)\lambda_{\text{max}}^2
\label{eq:extractability bound}
\end{equation}
where $\lambda_{\text{max}}$ is the Schimdt coefficient of the target state $\Psi'_{n}$. Combining Eqs. (\ref{eq:tradeoff}), (\ref{eq:liner beta}) and (\ref{eq:extractability bound}), the tradeoff function can be upper bounded by \cite{Kaniewski2016}
 \begin{equation}
     \mathcal{Q}\qty[{\Psi'_n,\mathscr{B},\beta_O}] \leq \lambda_{\text{max}}^2 + \qty(1-\lambda_{\text{max}}^2)\frac{\beta_O - \beta_L}{\beta_Q - \beta_L}
 \end{equation}
Note that, for the $n-$partite GHZ state, $\lambda_{max} = 1/{\sqrt{2}}$, thus for any Bell functional $\mathscr{B}$, we have 
\begin{equation}
    \mathcal{Q}\qty[{\Psi'_n,\mathscr{B},\beta_O}] \leq \frac{1}{2} + \frac{1}{2} \frac{\beta_O - \beta_L}{\beta_Q-\beta_L}
    \label{eq:tradeoff upper bound}
\end{equation}
The above expression holds for $n-$partite GHZ state for any Bell inequality for which $\beta_Q>\beta_L$. 
\subsection{Constructing the Operator inequality}
Our goal is to show that a large violation of considered Bell inequality implies that the underlying state has high fidelity with the optimal state $\Psi_n'$. This relation must be linear with the observed violation, thus holding for all possible measurements that can be performed by the parties involved in an experiment. To establish this, we first construct an operator inequality of the form ~\cite{Kaniewski2016}
\begin{equation}
    \mathcal{K}_n \succeq s\mathcal{W}_n + \mu \mathbb{I}
\label{eq:OPI}
\end{equation}
where $\mathcal{K}_n = \mathbf\Lambda^\dagger\Psi'_{n}$ is the state obtained after applying the adjoint of the extraction channel on the target state, $\mathcal{W}_n$ is the parametrised Bell operator constructed from $\mathscr{B}$, $\mathbb{I}$ is the $d \times d$ Identity matrix. Here, $s$ is a non-negative real constant and $\mu$ is a real constant chosen such that the inequality (\ref{eq:OPI}) is satisfied, then for a particular choice of $\Lambda$, the fidelity of any state $\rho_n$ with the optimal pure state $\Psi'_n$ is inner product
\begin{align}
    F(\mathbf{\Lambda}\rho_{n}, \Psi'_{n}) &= \Tr[\mathbf{\Lambda}\rho_n\Psi'_n] = \Tr[\rho_n\mathbf{\Lambda}^\dagger\Psi'_n] = \Tr[\mathcal{K}_n\ \rho_n]
\end{align}
where $\mathbf{\Lambda}^\dagger=\bigotimes_{i=1}^{n}\Lambda_{A_i}^\dagger$. Now taking trace on both sides of the inequality in Eq.~(\ref{eq:OPI}) with $\rho_n$ we have 
\begin{equation}
    F(\mathbf{\Lambda}\rho_{n}, \Psi'_{n}) \geq s \ \beta_O + \mu
\label{eq:f_beta}
\end{equation}
which constitutes a linear self-testing statement in terms of lower bounds on the fidelity of the underlying state with the ideal state as a function of observed Bell value $\beta_O$. 
Using Eqs.~(\ref{eq:extractibility}), (\ref{eq:tradeoff}) and (\ref{eq:f_beta}), we can derive a lower bound on the extractability--Bell violation tradeoff function
\begin{equation}
     \mathcal{Q}\qty[{\Psi'_n,\mathscr{B},\beta_O}] \geq s\,\beta_O + \mu.
     \label{eq:linear tradeoff}
\end{equation}
The function $\mathcal{Q}\qty[{\Psi'_n,\mathscr{B},\beta_O}]$ is constrained by the boundary conditions $\mathcal{Q}\qty[{\Psi'_n,\mathscr{B},\beta_T}] = 1/2$ for some threshold value $\beta_T \geq \beta_L$, beyond which the target state can be extracted with $\Xi \geq 1/2$. Additionaly, for optimal quantum bound $\beta_Q$, $\mathcal{Q}\qty[{\Psi'_n,\mathscr{B},\beta_Q}] = 1$. Since Eq.~(\ref{eq:linear tradeoff}) is linear in the observed violation $\beta_O$, these conditions uniquely determine the tradeoff relation. Consequently, the bound can be expressed as
\begin{equation}
    \mathcal{Q}\qty[{\Psi'_n,\mathscr{B},\beta_O}] 
    \geq \frac{1}{2} 
    + \frac{1}{2} \cdot \frac{\beta_O - \beta_T}{\beta_Q - \beta_T}
    \label{eq:tradeoff lower bound}
\end{equation}
As a consequence of Eqs. (\ref{eq:tradeoff upper bound}) and (\ref{eq:tradeoff lower bound}), it is easy to see that the STOPI method provides a tight bound on $\mathcal{Q}\qty[{\Psi'_n,\mathscr{B},\beta_O}]$ whenever the the threshold violation $\beta_T$ matches the local bound $\beta_L$, and a mismatch indicates towards the possible existence of a better robustness bound.
\section{Tools to be used: The qubit extraction map \label{Sec III}}
{In this work, we consider Svetlichny and MABK inequalities to self-test the GHZ state. For any arbitrary $n$, the maximization of both the functionals is achieved by $n-$qubit GHZ state and qubit observables due to Jordan lemma \cite{Panwar2023, Singh2025a}. Hence it is enough to consider local qubit extraction maps for the robustness analysis \cite{Kaniewski2016,Coopmans2019}. Without loss of generality, we can consider the local observables of each party to be projective, we consider the following measurements 
\begin{equation}
    A_j^{r}(\alpha_j) = \cos \alpha_j\, \sigma_x + (-1)^r \sin \alpha_j\, \sigma_y, \qquad r\in\{0,1\}
    \label{eq:LocalMeasurements}
\end{equation}
for $\alpha_j\in[0,\pi/2]$. 
{For the choice of local measurements in Eq. (\ref{eq:LocalMeasurements}) the Svetlichny and MABK operators from Eqs. (\ref{eq:SvOperator}) and (\ref{eq:MABK Operator}) can be respectively parameterised as
\begin{equation}
\mathcal{S}_n(\alpha_1\dots,\alpha_n)
=
\sum_{\mu=1}^{2^{n-1}}
\left[
\nu_\mu^{\,n}
\left(
A_0^{1}(\alpha_1)
+
(-1)^{n+1} A_1^{1}(\alpha_1)
\right)
\bigotimes_{j=2}^{n}
A^{x_{\mu,j}}_{j}(\alpha_j)
\right],
\label{eq:ParameterisedSvOperator}
\end{equation}
\noindent and,
\begin{equation}
\begin{aligned}
\mathcal{M}_n(\alpha_1\dots,\alpha_n)
=
&\left( \frac{1-i}{\sqrt{2}} \right)^{(n-1)\oplus 2}
\bigotimes_{j=1}^{n}
\left( A_j^{0}(\alpha_1) + i A_j^{1}(\alpha_1) \right) \\
&+
\left( \frac{1+i}{\sqrt{2}} \right)^{(n-1)\oplus 2}
\bigotimes_{j=1}^{n}
\left( A_j^{0}(\alpha_1) - i A_j^{1}(\alpha_1) \right)
\end{aligned}
\label{eq:Parameterised WMABK}
\end{equation}
{The corresponding $n-$ partite GHZ state can be obtained from the corresponding Bell operator as follows:
\begin{equation}
    \rho_{\mathcal{S}_n} = \frac{1}{2^n}\left(\bigotimes_{n}\mathbb{I} + \sum_i C_i +\eta_{\mathcal{S}}\left[\mathcal{S}_n\left(\frac{\pi}{4},...,\frac{\pi}{4}\right) \right]\right)
    \label{eq:RhoSn}
\end{equation}
and 
\begin{equation}
    \rho_{\mathcal{M}_n} = \frac{1}{2^n}\left(\bigotimes_{n}\mathbb{I} + \sum_i D_i +\eta_{\mathcal{M}}\left[\mathcal{M}_n\left(\frac{\pi}{4},...,\frac{\pi}{4}\right) \right]\right)
    \label{eq:RhoMn}
\end{equation}
where $C_i$ are the terms, each of which commutes with each term of $\mathcal{S}_n(\pi/4,...,\pi/4)$. The constant $\eta_\mathcal{S}$ is chosen such that $\Tr{\rho_{\mathcal{S}_n} \mathcal{S}_n(\pi/4,...,\pi/4)} = (\mathcal{S}_n)_Q$, similarly $D_i$ are the terms, each of which commutes with each term of $\mathcal{M}_n(\pi/4,...,\pi/4)$. The constant $\eta_\mathcal{M}$ is chosen such that $\Tr{\rho_{\mathcal{M}_n} \mathcal{M}_n(\pi/4,...,\pi/4)} = (\mathcal{M}_n)_Q$. We consider the parametrized  qubit dephasing channel of the extraction map is given by
\begin{equation}
\begin{aligned}
    K^{0}_j(\alpha_j) = \sqrt{\frac{1+g(\alpha_j)}{2}} \mathbb{I} \ ; \quad   K^{1}_j(\alpha_j) = \sqrt{\frac{1-g(\alpha_j)}{2}} \Gamma(\alpha_j)
\end{aligned}
\label{eq:DephasingChannel}
\end{equation}
where the dephasing parameter $g(\alpha_j)$ is defined as
\begin{equation}
    g(\alpha_j) = (1 + \sqrt{2}) (\sin{\alpha_j} + \cos{\alpha_j} - 1)
\label{eq:dephasing parameter}
\end{equation}
and 
\begin{equation}
    \begin{aligned}
    \Gamma(\alpha_j) &= \sigma_x &\alpha_j \in \left[0,\frac{\pi}{4}\right] \\
    &= \sigma_y & \alpha_j \in\left[\frac{\pi}{4},\frac{\pi}{2}\right]
    \end{aligned}
\label{eq:Gamma}
\end{equation}
where $j\in[n]$. The action of the channel (\ref{eq:DephasingChannel}) on an $n-$ partite state $\rho_{\mathcal{S}_n}$ and $\rho_{\mathcal{M}_n}$ are given by \cite{Wilde2017}
\begin{equation}
    \mathcal{K}_{\mathcal{S}_n}(\alpha_1, \dots, \alpha_n) = \sum_{r_1\dots r_j\in\{0,1\}}\left(\bigotimes_{j=1}^nK_j^{r_j}(\alpha_j)\right)\rho_{\mathcal{S}_n}\left(\bigotimes_{j=1}^nK_j^{r_j}(\alpha_j)\right)^\dagger
\end{equation}
and
\begin{equation}
    \mathcal{K}_{\mathcal{M}_n}(\alpha_1, \dots, \alpha_n) = \sum_{r_1\dots r_j\in\{0,1\}}\left(\bigotimes_{j=1}^nK_j^{r_j}(\alpha_j)\right)\rho_{\mathcal{M}_n}\left(\bigotimes_{j=1}^nK_j^{r_j}(\alpha_j)\right)^\dagger
\end{equation}
For this choice of the extraction channel, it is clear that $g(\alpha_j) = g(\pi/2-\alpha_j)$; which means that the amount of dephasing of $\rho_n$ for any $\alpha_j$ is the same as the dephasing by $(\pi/2-\alpha_j)$. Additionally, the dephasing parameters are symmetric about $\alpha_j=\pi/4$, so qualitatively, any result valid for $\alpha_j \in [0,\pi/4]$ will also be valid for $\alpha_j \in [0,\pi/2]$ hence we only need to prove our results in the domain $\alpha_j \in [0,\pi/4]$.}

\section{Robustness of Svetlichny based self-testing protocol \label{Sec IV}}
Let the observed value of the Svetlichny functional be denoted by $(\mathcal{S}_n)_O$. Since this value must exceed the local bound, we require $(\mathcal{S}_n)_O \geq (\mathcal{S}_n)_L$. Defining $\qty(\mathcal{S}_n)_T \geq (\mathcal{S}_n)_L$ as the threshold value of the Svetlichny operator above which robust self-testing becomes achievable. In other words, once this threshold is surpassed, the fidelity between the shared state among the parties and the $n$-partite GHZ state exceeds 0.5. This is captured in the following theorem,
{\thm If the observed value of the Svetlichny functional $(\mathcal{S}_n)_O$ is greater than the threshold value $\qty(\mathcal{S}_n)_T$, then there exists a local extraction channel $\Lambda$ with $\mathcal{K}_{\mathcal{S}_n}=\Lambda^\dag\qty(\rho_{\mathcal{S}_n})$, such that the following operator inequality 
\begin{equation}
    \mathcal{K}_{\mathcal{S}_n} \succeq s\mathcal{S}_n + \mu \mathbb{I}
    \label{eq: Sv Operator Inequality}
\end{equation}
holds for 
\begin{equation}
    \begin{aligned}
        i)\ n&=3, \quad s = \frac{3}{16}\left(1+\sqrt{2}\right), \quad \mu = -\frac{2+3\sqrt{2}}{4}, \\ &\qquad \qty(\mathcal{S}_3)_T = \frac{4}{3}\qty(2+\sqrt{2})\\
        ii)\ n&=4, \quad s = \frac{1}{16}\left(1+\sqrt{2}\right), \quad\mu = -\frac{1}{\sqrt{2}}, \\ &\qquad\qty(\mathcal{S}_4)_T = (\mathcal{S}_4)_L = 8\\
        iii)\ n&=5, \quad s = \frac{1}{32}\left(1+\sqrt{2}\right), \quad\mu = -\frac{1}{\sqrt{2}}, \\ &\qquad\qty(\mathcal{S}_5)_T = (\mathcal{S}_5)_L= 16
    \end{aligned}
\end{equation}
and the extraction channel in Eq.~(\ref{eq:DephasingChannel}) extracts a state with the extractibility 
\begin{equation}\label{eq:SvetlichnyQ}
    \mathcal{Q}\qty[\rho_{\mathcal{S}_n},\mathcal{S}_n,\qty(\mathcal{S}_n)_O] = \frac{1}{2} + \frac{1}{2}\frac{\qty(\mathcal{S}_n)_O - \qty(\mathcal{S}_n)_L}{(\mathcal{S}_n)_{Q} - (\mathcal{S}_n)_{L}}
\end{equation} 

\begin{proof}
    The proof is quite lengthy and hence deferred to Appendices \ref{svet3} and \ref{svet4}. Here we provide a sketch of it. Since each of the parties has a choice between two dichotomic measurements, using the Jordan lemma, we reduce the problem to the Hilbert space of dimension two. The measurements in Eq.~(\ref{eq:LocalMeasurements}) are antidiagonal matrices. Note that, tensor product of antidiagonal matrices is also an antidiagonal matrix. Now, $\mathcal{S}_n(\alpha_1, \dots, \alpha_n)$ is an antidiagonal matrix because it is a linear combination of antidiagonal matrices. Now, the terms $C_i$ in Eq.~(\ref{eq:RhoSn}) are antidiagonals if each of them commutes with terms in $\mathcal{S}_n\qty(\frac{\pi}{4},\dots,\frac{\pi}{4})$ which are antidiagonal. Hence, the matrix $\rho_{\mathcal{S}_n}$ is a persymmetric matrix since it must be hermitian and is formed by a linear combination of diagonal $\mathbb{I}$ and antidiagonal matrices. As shown in Appx.~\ref{persymmetricityproof}, $\mathcal{K}(\alpha_1, \dots, \alpha_n)$ is also a persymmetrix matrix. Thus the matrix
    \begin{equation}
        T_n=\mathcal{K}_{\mathcal{S}_n}(\alpha_1,\alpha_2,\dots,\alpha_n)-s\mathcal{S}_n(\alpha_1,\alpha_2,\dots,\alpha_n)-\mu\mathbb{I}
    \end{equation}
    is a Hermitian matrix which is a linear combination of a persymmetric matrix, an antidiagonal matrix and a diagonal matrix. Hence, $T_n$ is a sparse persymmetric matrix, and by using a unitary matrix
    \begin{equation}
        U = \sum_{k=0}^{2^{n-1}-1}\ket{2k}\bra{k}+\sum_{k=2^{n-1}}^{2^{n}-1}\ket{2^{n+1}-1-2k}\bra{k}
    \end{equation} 
for any $n$, the matrix $T_n$ can be brought into a block diagonal form using $U$. The block diagonal form of $T_n $ contains $2^{n-1}$ number of $2\times2$ blocks. Positivity of each of the block establishes the positivity of $T_n$ which proves the theorem statements. A detailed proof using this method for i) and ii) can be found in Appx.~\ref{svet3}, and Appx.~\ref{svet4} respectively. As a consequence of Eqs. (\ref{eq:tradeoff upper bound}) and (\ref{eq:tradeoff lower bound}), it is straightforward to conclude that for $n=4,5$, the threshold violation $\qty(\mathcal{S}_n)_T =\qty(\mathcal{S}_n)_L$, hence the STOPI method provides a tight bound on $\mathcal{Q}\qty[{\Psi'_n,\mathcal{S}_n,\qty(\mathcal{S}_n)_L}]$ resulting Eq.~(\ref{eq:SvetlichnyQ}).
\end{proof}
}
Note that for a non-genuine correlation generated by $(n-1)$-separable state, the value of the Svetlichny functional would be the local bound $\qty(\mathcal{S}_n)_L$. This is due to the fact that any violation of Svetlichny inequality detects a genuinely nonlocal correlation\cite{Svetlichny1987, Seevinck2002}. Now, the equality in Eq.~(\ref{eq:SvetlichnyQ}) holds for $n=4,5$. For $n=3$, our results show that $\qty(\mathcal{S}_3)_T\neq\qty(\mathcal{S}_3)_L$, however, due to the results in $n=4,5$ cases, we believe that there exists an extraction channel for which $\qty(\mathcal{S}_3)_T=\qty(\mathcal{S}_3)_L$. 
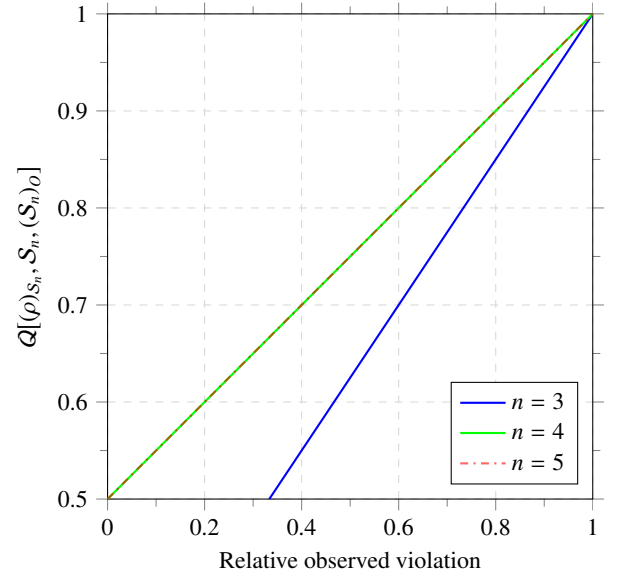
\begin{figure}[h]
  \centering
  \begin{tikzpicture}
    \begin{axis}[
        width  = 8cm,
        height = 8cm,
        xlabel = {Relative observed violation},
        ylabel = {$\mathcal{Q}\qty[(\rho)_{\mathcal{S}_n},\mathcal{S}_n,\qty(\mathcal{S}_n)_O]$},
        xmin = 0,  xmax = 1,
        ymin = 0.5, ymax = 1,
        legend pos  = south east,
        legend style = {font=\small},
        grid = major,
        grid style = {dashed, gray!30},
        tick align = outside,
        minor tick num = 1,
      ]

      \addplot[
        color = blue,
        thick,
      ] coordinates {
        (0.333333, 0.500000)
        (1.000000, 1.000000)
      };
      \addlegendentry{$n=3$}

      \addplot[ color = green,
      thick,
      ]
      coordinates {
      (0.000000, 0.500000)
      (1.000000, 1.000000) 
      };
      \addlegendentry{$n=4$}

      \addplot[
        color = red,
        thick,
        dashdotted,
        opacity = 0.6,
      ] coordinates {
        (0.000000, 0.500000)
        (1.000000, 1.000000)
      };
      \addlegendentry{$n=5$}

    \end{axis}
  \end{tikzpicture}
  \caption{$\mathcal{Q}\qty[\rho_{\mathcal{S}_n},\mathcal{S}_n,\qty(\mathcal{S}_n)_O]$ as a function of relative observed violation of the Svetlichny inequality for $n = 3, 4, 5$.}
  \label{fig:sv}
\end{figure}
\section{Robustness of MABK Based Protocol \label{Sec V}}
Analogous to the Svetlichny-based protocol, we denote by $(\mathcal{M}_n)_O$ and $(\mathcal{M}_n)_T$ the observed value and the threshold value of the MABK operator, respectively, such that $(\mathcal{M}_n)_O \geq (\mathcal{M}_n)_L$ and $(\mathcal{M}_n)_T \geq (\mathcal{M}_n)_L$. Note that for $n=3$, MABK inequality reduces to Mermin inequality. In \cite{Kaniewski2016}, it was shown that the inequality  
\begin{equation}
    \mathcal{K}_{\mathcal{M}_3} \succeq s\mathcal{M}_3 + \mu \mathbb{I}
\end{equation}
holds for $s = \frac{1}{8}\qty(2+\sqrt{2})$ and $\mu = -\frac{1}{\sqrt{2}}$. Hence, a state with fidelity greater than half with the $3-$partite GHZ state can be extracted from the suboptimal state shared by the three parties, which gives the Mermin value of $(\mathcal{M}_3)_O$. The threshold value $(\mathcal{M}_3)_T = 2\sqrt{2}$ was found to be trivial since it is the maximal value achievable by a state equivalent to the state of the form $\ket{\phi^+}_{A_1,A_2}\otimes\ket{0}_{A_3}$. {This state gives a fidelity of half with $3-$partite GHZ while producing the maximum possible Mermin value. 

Similarly, to analyze the bounds in the scenarios with $n$ parties, we must consider a state of the form $\sigma_n = \rho_{\mathcal{M}_{n-1}}\otimes\ket{0}\bra{0}$, which means that $(n-1)$ parties receive one particle each from $\rho_{\mathcal{M}_{n-1}}$ (\emph{i.e.,} $(n-1)$ GHZ state) and one party receives $\ket{0}\bra{0}$. Notice, $\sigma_n$ has a fidelity of half with $\rho_{\mathcal{M}_{n}}$. It is easy to see that this state produces maximum MABK value when $n-1$ parties implement local anticommuting observables and one party implements only one measurement.} Now, we state the following theorem for $n=4$ and $5$.
{\thm If the observed value of the MABK functional $(\mathcal{M}_n)_O$ is greater than the threshold value $\qty(\mathcal{M}_n)_T$, then there exists a local extraction channel $\Lambda$ with $\mathcal{K}_{\mathcal{M}_n}=\Lambda^\dag\qty(\rho_{\mathcal{M}_n})$, such that the following operator inequality 
\begin{equation}\label{opineqmermin}
    \mathcal{K}_{\mathcal{M}_n} \succeq s\mathcal{M}_n + \mu \mathbb{I} 
\end{equation}
holds for 
\begin{equation}
    \begin{aligned}
        i)\ n&=4, \quad s = \frac{1}{16}\left(2+\sqrt{2}\right), \quad\mu = -\frac{1}{\sqrt{2}}, \\ &\qquad\qty(\mathcal{M}_4)_T = 4\sqrt{2}\\
        ii)\ n&=5, \quad s = \frac{1}{32}\left(2+\sqrt{2}\right), \quad\mu = -\frac{1}{\sqrt{2}}, \\ &\qquad\qty(\mathcal{M}_5)_T = 8\sqrt{2}
    \end{aligned}
\end{equation}
and the extraction channel in Eq.~(\ref{eq:DephasingChannel}) extracts a state with the extractability
\begin{equation}
    \mathcal{Q}\qty[\rho_{\mathcal{M}_n},\mathcal{M}_n,\qty(\mathcal{M}_n)_O] = \frac{1}{2} + \frac{1}{2}\frac{\qty(\mathcal{M}_n)_O - \qty(\mathcal{M}_n)_T}{(\mathcal{M}_n)_{Q} - (\mathcal{M}_n)_{T}}
\label{eq:MABK Tradeoff}
\end{equation}
}
\begin{proof}
  Proof of Eq.~(\ref{opineqmermin}) for i) and ii) is similar to proof in Theorem 1.  By using $\sigma_n$ in  Eq. (\ref{eq:State Equation}) for $n-4,5$, we obtain the upper bound on $\mathcal{Q}$ given by
\begin{equation}
    \mathcal{Q}\qty[\rho_{\mathcal{M}_n},\mathcal{M}_n,\qty(\mathcal{M}_n)_O] \leq \frac{1}{2} + \frac{1}{2}\frac{\qty(\mathcal{M}_n)_O - \qty(\mathcal{M}_n)^*}{(\mathcal{M}_n)_{Q} - (\mathcal{M}_n)^{*}}
\label{eq:MABK bipartite cut}
\end{equation}
where $(\mathcal{M}_n)^*$ is $4\sqrt{2}$ and $8\sqrt{2}$ when the number of parties is four and five respectively. Now, for $n=4,5$ $(\mathcal{M}_n)^*=(\mathcal{M}_n)_T$. So, the lower bound (\ref{eq:MABK Tradeoff}) on $\mathcal{Q}$  matches exactly with the upper bound, (\ref{eq:MABK bipartite cut}), and hence the extractability of the extracted state with the ideal state for the protocol to have non-trivial fidelity as expressed in Eq. (\ref{eq:MABK Tradeoff}).
\end{proof}
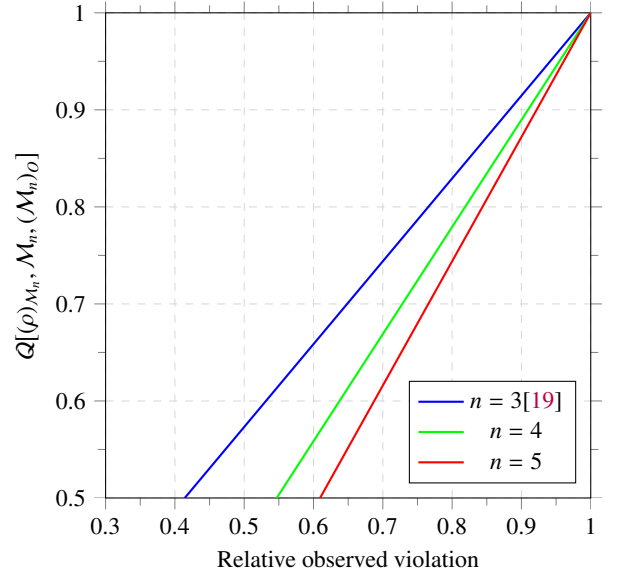
\begin{figure}[ht]
  \centering
  \begin{tikzpicture}
    \begin{axis}[
        width  = 8cm,
        height = 8cm,
        xlabel = {Relative observed violation},
        ylabel = {$\mathcal{Q}\qty[(\rho)_{\mathcal{M}_n},\mathcal{M}_n,\qty(\mathcal{M}_n)_O]$},
        xmin = 0.3, xmax = 1,
        ymin = 0.5, ymax = 1,
        legend pos  = south east,
        legend style = {font=\small},
        grid = major,
        grid style = {dashed, gray!30},
        tick align = outside,
        minor tick num = 1,
      ]

      \addplot[
        color = blue,
        thick,
      ] coordinates {
        (0.414213, 0.500000)
        (1.000000, 1.000000)
      };
      \addlegendentry{$n=3$\cite{Kaniewski2016}}

      \addplot[
        color = green,
        thick,
      ] coordinates {
        (0.546918, 0.500000)
        (1.000000, 1.000000)
      };
      \addlegendentry{$n=4$}

      \addplot[
        color = red,
        thick,
      ] coordinates {
        (0.609475, 0.500000)
        (1.000000, 1.000000)
      };
      \addlegendentry{$n=5$}

    \end{axis}
  \end{tikzpicture}
  \caption{$\mathcal{Q}\qty[\rho_{\mathcal{M}_n},\mathcal{M}_n,\qty(\mathcal{M}_n)_O]$ as a function of relative observed violation of the MABK inequality for $n = 3, 4, 5$.}
  \label{fig:mabk}
\end{figure}
 Since, for the MABK operator, the violation of the local bound does not imply nontrivial extraction of a GHZ state, self-testing schemes based on MABK are not experimentally friendly. Nontrivial extractability of GHZ requires a large value of the MABK operator, which may be difficult to attain experimentally. This is evident from the graph in Fig.~\ref{fig:mabk}: the relative violation required for nontrivial extractability increases as the number of parties increases. In contrast, the Svetlichny inequality for any number of parties requires the violation of the local bound to establish non-trivial extractability with the GHZ state.  

\section{Discussion and Future Work}
{In this work, we have analysed the robustness of self-testing protocols for the $n-$partite GHZ state based on Svetlichny and MABK inequalities. We found that both of the self-testing schemes have trivial lower bound for extractibility-bell violation function for $n=4,5$. We obtained a nontrivial bound of $\qty(\mathcal{S}_3)_T\approx4.55$ for $n=3$, however, we strongly believe that by using alternative extraction channels \cite{Chen2026}, one can obtain trivial robust self-testing bound for the $3-$partite Svetlichny inequality. Although both the Svetlichny and the MABK inequality based self-testing schemes have trivial lower bounds for robustness, we demonstrated that the Svetlichny based scheme is better suited for robust self-testing of GHZ state because any value $(\mathcal{S}_n)_O > (\mathcal{S}_n)_L$ establishes a robust self-test of the GHZ state. This is in contrast to self-testing based on the MABK operator which requires very large values of observed violation as the number of parties increase, making the MABK based self-testing approach infeasible for certifying GHZ states shared between large number of parties. 

When the choice of local observables is two, Bell operators proposed by Sarkar \textit{et.al.} \cite{Sarkar2022} reduces to Svetlichny operator if we take the local dimensions of the system to be self-tested as two. The self-testing scheme of Baccari \textit{et. al.} \cite{Baccari2020} for self-testing GHZ is comparable in it's robustness to the MABK based protocol since both of them yield similar lower bounds for fidelities with the $n$-partite GHZ for $n=3,4$ and $5$. Similar to the MABK, Baccari \textit{et. al.} \cite{Baccari2020} protocol is not suited for self-testing GHZ states when the number of parties is large. For example, for $n=5$, about $68\%$ of relative violation is required to extract a state of fidelity greater than half with the $5-$partite GHZ state Which is similar to the MABK based protocol as shown in Fig.~\ref{fig:mabk}. On the other hand, our analytical results show that when the number of parties is greater than three, the Svetlichny inequality based protocol extracts a state with fidelity greater than half with the GHZ state for an arbitrarily small violation of the Svetlichny inequality. We conjecture that the pattern found for $n=4$ and $5$ in the Svetlichny based protocol remains the same for arbitrary $n>5$ parties. 

We have examined the robustness of self-testing schemes in multipartite scenarios where each party performs two dichotomic measurements, and have presented a fully analytical approach to robustly self-test the GHZ state. In these scenarios, Jordan’s lemma applies, enabling us to restrict the analysis to the case in which the measurements, the shared state, and the extraction channels all have local dimension of two. It would then be interesting to extend our approach to the bipartite  and multipartite scenarios beyond two-input and two-output scenarios. A notable bipartite example is the Elegant Bell inequality \cite{Gisin2007}, and its generalizations \cite{Ghorai2018} and multi input chained Bell inequalities \cite{Paul2026}. Analytical robustness quantification techniques could also be extended to Bell inequalities that self-test bipartite qudit entangled states \cite{Coladangelo2017, Coladangelo2017b, Sarkar2021} as well as multipartite qudit entangled states \cite{Sarkar2022}.  

\section{Acknowledgements}
PKJ acknowledges funding from the University Grants Commission (NTA Ref. No.- 231620086501), Government of India . RKS acknowledges the financial support from the Council of Scientific and Industrial Research (CSIR, 09/1001(17051)/2023-EMR-I), Government of India.  AKP acknowledges the support from the Research Grant SERB/CRG/2021/004258, Government of India.
\bibliography{references} 
\appendix
\begin{widetext}
\section{Proof that $\mathcal{K}_{\mathcal{S}_n}\qty(\alpha_1,\dots,\alpha_n)$  is persymmetric}\label{persymmetricityproof}
For a matrix $\mathscr{M} \in  \mathcal{L}(\mathcal{H})$ to be persymmetric, $\mathscr{M} = J \mathscr{M}^T J$, where $J$ is the exchange matrix. Since $\rho_{\mathcal{S}_n}$ is persymmetric, we have
\begin{equation}
    \rho_{\mathcal{S}_n} = J \rho_{\mathcal{S}_n}^T J \label{eq:rho_persym}
\end{equation}
Now, the channel action on $\rho_{\mathcal{S}_n}$ yields
\begin{equation}
    \mathcal{K}_{\mathcal{S}_n} = \sum_{r_1 \dots r_n \in \{0,1\}} \left( \bigotimes_{j=1}^n K_j^{r_j}(\alpha_j) \right) \rho_{\mathcal{S}_n} \left( \bigotimes_{j=1}^n K_j^{r_j}(\alpha_j) \right)
\end{equation}
Using Eq. (\ref{eq:rho_persym}) in the above equation,
\begin{equation}
    \mathcal{K}_{\mathcal{S}_n} = \sum_{r_1 \dots r_n \in \{0,1\}} \left( \bigotimes_{j=1}^n K_j^{r_j}(\alpha_j) \right) J \rho_{\mathcal{S}_n}^T J \left( \bigotimes_{j=1}^n K_j^{r_j}(\alpha_j) \right)
\end{equation}
If $\mathcal{K}_{\mathcal{S}_n}$ is persymmetric, then $\mathcal{K}_{\mathcal{S}_n} = J \mathcal{K}_{\mathcal{S}_n}^T J$. To prove this, we evaluate 
\begin{align}
    J \mathcal{K}_{\mathcal{S}_n}^T J &= J \sum_{r_1 \dots r_n} \left( \bigotimes_{j=1}^n K_j^{r_j}(\alpha_j) \right)^T J \rho_{\mathcal{S}_n} J \left( \bigotimes_{j=1}^n K_j^{r_j}(\alpha_j) \right)^T J \nonumber \\
    &= \sum_{r_1 \dots r_n} \left( J \bigotimes_{j=1}^n \left[K_j^{r_j}(\alpha_j)\right]^T J \right) \rho_{\mathcal{S}_n} \left( J \bigotimes_{j=1}^n \left[K_j^{r_j}(\alpha_j)\right]^T J \right)
\end{align}
In Pauli basis, the exchange matrix $J = \bigotimes_{j=1}^n \sigma_x$, thus we have (with simplified notation $g(x)\equiv g_x$)
\begin{equation}
    J \mathcal{K}_{\mathcal{S}_n}^T J = \sum_{r_1 \dots r_n} \left( \bigotimes_{j=1}^n \sigma_x \left[K_j^{r_j}(\alpha_j)\right]^T \sigma_x \right) \rho_{\mathcal{S}_n} \left( \bigotimes_{j=1}^n \sigma_x \left[K_j^{r_j}(\alpha_j)\right]^T \sigma_x \right)
\end{equation}
now, the Kraus operators as defined in Eqs.~(\ref{eq:DephasingChannel})-(\ref{eq:Gamma}) are
\begin{equation}
    K_j^{r_j}(\alpha_j) = 
    \begin{cases} 
        \sqrt{\frac{1+g_{\alpha_j}}{2}} \mathbb{I} & \text{if } r_j = 0 \\
        \sqrt{\frac{1-g_{\alpha_j}}{2}} \sigma_x & \text{if } r_j = 1, \, \alpha_j \in \left[0, \frac{\pi}{4}\right] \\
        \sqrt{\frac{1-g_{\alpha_j}}{2}} \sigma_y & \text{if } r_j = 1, \, \alpha_j \in \left(\frac{\pi}{4}, \frac{\pi}{2}\right] 
    \end{cases}
\end{equation}
We evaluate $\sigma_x \left[K_j^0(\alpha_j)\right]^T \sigma_x$ for each case. Firstly, for $r_j = 0$:
\begin{equation}
    \sigma_x \left[K_j^0(\alpha_j)\right]^T \sigma_x = K_j^0(\alpha_j)
\end{equation}
For $r_j = 1$ and $\alpha_j \in \left[0, \frac{\pi}{4}\right]$:
\begin{align}
    K_j^{r_j} &= \sqrt{\frac{1-g_{\alpha_j}}{2}} \sigma_x  \implies 
    (K_j^{r_j})^T = \sqrt{\frac{1-g_{\alpha_j}}{2}} \sigma_x = K_j^{r_j} \quad \text{Thus} \ \  
    \sigma_x K_j^{r_j} \sigma_x = K_j^{r_j}
\end{align}
For $r_j = 1$ and $\alpha_j \in \left(\frac{\pi}{4}, \frac{\pi}{2}\right]$
\begin{align}
    K_j^{r_j} &= \sqrt{\frac{1-g_{\alpha_j}}{2}} \sigma_y \implies (K_j^{r_j})^T = -\sqrt{\frac{1-g_{\alpha_j}}{2}} \sigma_y \nonumber \\
    \sigma_x (K_j^{r_j})^T \sigma_x &= -\sqrt{\frac{1-g_{\alpha_j}}{2}} \sigma_x \sigma_y \sigma_x = \sqrt{\frac{1-g_{\alpha_j}}{2}} \sigma_y = K_j^{r_j}
\end{align}
Therefore, $\sigma_x \left[K_j^{r_j}(\alpha_j)\right]^T \sigma_x = K_j^{r_j}(\alpha_j) \ \forall r_j, \alpha_j$. Thus,
\begin{equation}
    J \mathcal{K}_{\mathcal{S}_n}^T J = \mathcal{K}_{\mathcal{S}_n}
\end{equation}
Which proves that $\mathcal{K}_{\mathcal{S}_n}$ is persymmetric.
\section{Robustness of the self-testing protocol based on 3-Partite Svetlichny inequality}\label{svet3}
\noindent For $n = 3$, by using Eq.~(\ref{eq:LocalMeasurements}) the Svetlichny functional in Eq.~(\ref{eq:ParameterisedSvOperator}) can be written as
\begin{equation}
\begin{aligned}
\mathcal{S}_3(\alpha_1,\alpha_2,\alpha_3)=4\Big(
&-\cos(\alpha_1)\cos(\alpha_2)\cos(\alpha_3)\, \sigma_x\otimes\sigma_x\otimes\sigma_x
+\cos(\alpha_1)\sin(\alpha_2)\sin(\alpha_3)\, \sigma_x\otimes\sigma_y\otimes\sigma_y \\
&+\sin(\alpha_1)\cos(\alpha_2)\sin(\alpha_3)\, \sigma_y\otimes\sigma_x\otimes\sigma_y
+\sin(\alpha_1)\sin(\alpha_2)\cos(\alpha_3)\, \sigma_y\otimes\sigma_y\otimes\sigma_x
\Big)
\label{eq:W3Sv}
\end{aligned}
\end{equation}
The $3-$partite GHZ state which gives the maximum value $(\mathcal{S}_3)_Q = 4\sqrt{2}$ as in Eq.~(\ref{eq:RhoSn}) is given by
\begin{equation}
\begin{aligned}
\rho_{\mathcal{S}_3} = \frac{1}{8} \Big[
    &\mathbb{I} \otimes \mathbb{I} \otimes \mathbb{I}
    + \sigma_z \otimes \sigma_z \otimes \mathbb{I}
    + \mathbb{I} \otimes \sigma_z \otimes \sigma_z
    + \sigma_z \otimes \mathbb{I} \otimes \sigma_z
    \\
    &- \sigma_x \otimes \sigma_x \otimes \sigma_x
    + \sigma_x \otimes \sigma_y \otimes \sigma_y
    + \sigma_y \otimes \sigma_x \otimes \sigma_y
    + \sigma_y \otimes \sigma_y \otimes \sigma_x
\Big]
\end{aligned}
\label{3svs}
\end{equation}
After applying the dephasing channel defined in Eq.~(\ref{eq:DephasingChannel}), the state (\ref{3svs}) becomes
\begin{equation}
\begin{aligned}
\mathcal{K}_{\mathcal{S}_3}(\alpha_1,\alpha_2,\alpha_3)
= \frac{1}{8}\Bigl[
    &\mathbb{I}\otimes\mathbb{I}\otimes\mathbb{I}
    + g_{\alpha_1}g_{\alpha_2}\,\sigma_z\otimes\sigma_z\otimes\mathbb{I}
    + g_{\alpha_2}g_{\alpha_3}\,\mathbb{I}\otimes\sigma_z\otimes\sigma_z \\[4pt]
    &+ g_{\alpha_1}g_{\alpha_3}\,\sigma_z\otimes\mathbb{I}\otimes\sigma_z
    - \sigma_x\otimes\sigma_x\otimes\sigma_x
    + g_{\alpha_2}g_{\alpha_3}\,\sigma_x\otimes\sigma_y\otimes\sigma_y \\[4pt]
    &+ g_{\alpha_1}g_{\alpha_3}\,\sigma_y\otimes\sigma_x\otimes\sigma_y
    + g_{\alpha_1}g_{\alpha_2}\,\sigma_y\otimes\sigma_y\otimes\sigma_x
\Bigr]
\end{aligned}
\label{eq:3svK}
\end{equation}
To prove the robustness claim, we need to prove that $\mathcal{K}_{\mathcal{S}_3}(\alpha_1,\alpha_2,\alpha_3)\succeq s\mathcal{S}_3(\alpha_1,\alpha_2,\alpha_3)-\mu\mathbb{I}\otimes\mathbb{I}\otimes\mathbb{I}$, which implies that we have to prove $T_3=\mathcal{K}_{\mathcal{S}_3}(\alpha_1,\alpha_2,\alpha_3)-s\mathcal{S}_3(\alpha_1,\alpha_2,\alpha_3)-\mu\mathbb{I}\otimes\mathbb{I}\otimes\mathbb{I}\succeq0$. Now, the operator, $T_3 $ is a matrix which has diagonal and antidiagonal entries. We consider a unitary matrix of the following form
\begin{equation}
    U = \sum_{k=0}^3 \ket{2k}\bra{k} +\sum_{k=4}^{7} \ket{15-2k}\bra{k}
\end{equation}
which gives $UT_3U^\dag = Q_1 \oplus Q_2\oplus Q_3 \oplus Q_4$ where,
\begin{equation}
\begin{aligned}
Q_1 &= \begin{bmatrix}
f_1(\alpha_1,\alpha_2,\alpha_3,s) & f_2(\alpha_1,\alpha_2,\alpha_3,s)\\
f_2(\alpha_1,\alpha_2,\alpha_3,s) & f_1(\alpha_1,\alpha_2,\alpha_3,s)
\end{bmatrix};
\quad
Q_2 = \begin{bmatrix}
f_3(\alpha_1,\alpha_2,\alpha_3,s) & f_4(\alpha_1,\alpha_2,\alpha_3,s)\\
f_4(\alpha_1,\alpha_2,\alpha_3,s) & f_3(\alpha_1,\alpha_2,\alpha_3,s)
\end{bmatrix}
\\
Q_3 &= \begin{bmatrix}
f_5(\alpha_1,\alpha_2,\alpha_3,s) & f_6(\alpha_1,\alpha_2,\alpha_3,s)\\
f_6(\alpha_1,\alpha_2,\alpha_3,s) & f_5(\alpha_1,\alpha_2,\alpha_3,s)
\end{bmatrix};
\quad
Q_4 = \begin{bmatrix}
f_7(\alpha_1,\alpha_2,\alpha_3,s) & f_8(\alpha_1,\alpha_2,\alpha_3,s)\\
f_8(\alpha_1,\alpha_2,\alpha_3,s) & f_7(\alpha_1,\alpha_2,\alpha_3,s)
\end{bmatrix}
\end{aligned}
\end{equation}
where the functions inside the matrices are defined as 
\begin{equation}
    \begin{aligned}
        f_1(\alpha_1,\alpha_2,\alpha_3,s) &= \frac{1}{8}\qty[-7 + g_{\alpha_2} g_{\alpha_3} + g_{\alpha_1}(g_{\alpha_2} + g_{\alpha_3})] + 4\sqrt{2}s\\
        f_2(\alpha_1,\alpha_2,\alpha_3,s) &= \frac{1}{8}\qty[-1 - g_{\alpha_2} g_{\alpha_3} - g_{\alpha_1}(g_{\alpha_2} + g_{\alpha_3})]+ 4s \cos(\alpha_1-\alpha_2)\cos(\alpha_3) +4s\sin(\alpha_1+\alpha_2)\sin(\alpha_3)\\
        f_3(\alpha_1,\alpha_2,\alpha_3,s) &= \frac{1}{8}\qty[-7 + g_{\alpha_1} g_{\alpha_2} - (g_{\alpha_1} + g_{\alpha_2}) g_{\alpha_3}] + 4\sqrt{2}\,s\\
        f_4(\alpha_1,\alpha_2,\alpha_3,s) &= \frac{1}{8}\qty[-1 - g_{\alpha_1} g_{\alpha_2} + (g_{\alpha_1} + g_{\alpha_2}) g_{\alpha_3}]
+ 4s \cos(\alpha_1-\alpha_2)\cos(\alpha_3) - 4s \sin(\alpha_1+\alpha_2)\sin(\alpha_3)\\
        f_5(\alpha_1,\alpha_2,\alpha_3,s) &= \frac{1}{8}\qty[-7 - g_{\alpha_2} g_{\alpha_3} + g_{\alpha_1}(-g_{\alpha_2} + g_{\alpha_3})] + 4\sqrt{2}s\\
        f_6(\alpha_1,\alpha_2,\alpha_3,s) &= \frac{1}{8}\qty[-1 + g_{\alpha_1}(g_{\alpha_2} - g_{\alpha_3}) + g_{\alpha_2} g_{\alpha_3}]
+ 4s \cos(\alpha_1 + \alpha_2)\cos(\alpha_3) + 4s \sin(\alpha_1 - \alpha_2)\sin(\alpha_3)\\
        f_7(\alpha_1,\alpha_2,\alpha_3,s) &= \frac{1}{8}\qty[-7 + g_{\alpha_2} g_{\alpha_3} - g_{\alpha_1}(g_{\alpha_2} + g_{\alpha_3})] + 4\sqrt{2}s\\
        f_8(\alpha_1,\alpha_2,\alpha_3,s) &= \frac{1}{8}\qty[-1 - g_{\alpha_2} g_{\alpha_3} + g_{\alpha_1}(g_{\alpha_2} + g_{\alpha_3})]
+ 4s \cos(\alpha_1 + \alpha_2)\cos(\alpha_3) - 4s \sin(\alpha_1 - \alpha_2)\sin(\alpha_3)
    \end{aligned}
\end{equation}
Now, to prove positivity of $T_3$, we need to show that each of the $Q_i\succeq0$.
\subsubsection{Positivity of $Q_1$}
To show that $Q_1\succeq0$, we will show that both of it's eigenvalues are greater than or equal to zero. The two eigenvalues of $Q_1$ are simply $\lambda_{\pm}(\alpha_1,\alpha_2,\alpha_3,s) = f_1(\alpha_1,\alpha_2,\alpha_3,s)\pm f_2(\alpha_1,\alpha_2,\alpha_3,s)$. Considering the first eigenvalue,
\begin{equation}
    \begin{aligned}
        \lambda_{+}(\alpha_1,\alpha_2,\alpha_3,s) &= f_1(\alpha_1,\alpha_2,\alpha_3,s)+ f_2(\alpha_1,\alpha_2,\alpha_3,s) = -1 + 4s\qty(\sqrt{2} + \cos(\alpha_1-\alpha_2)\cos(\alpha_3) + \sin(\alpha_1+\alpha_2)\sin(\alpha_3))
    \end{aligned}
\end{equation}
The function $h_1(\alpha_1,\alpha_2,\alpha_3) = \cos(\alpha_1-\alpha_2)\cos(\alpha_3) + \sin(\alpha_1+\alpha_2)\sin(\alpha_3)$ is minimized at the corners of the cube $\alpha_1,\alpha_2,\alpha_3\in[0,\frac{\pi}{4}]$ because in this domain, it's gradient is never zero. Then, $\min(h_1(\alpha_1,\alpha_2,\alpha_3)) = \frac{1}{\sqrt{2}}$ which implies that $\min (\lambda_+) = -1+6\sqrt{2}s$ which is greater than zero for $s=\frac{3\qty(1+\sqrt{2})}{16}$.

Now, consider the second eigenvalue $\lambda_{-} = f_1(\alpha_1,\alpha_2,\alpha_3,s) - f_2(\alpha_1,\alpha_2,\alpha_3,s)$, which can be explicitly written as
\begin{equation}
    \begin{aligned}
        \lambda_{-}(\alpha_1,\alpha_2,\alpha_3,s) = \frac{1}{4}\qty(-3 + g_{\alpha_1}g_{\alpha_2} + g_{\alpha_1}g_{\alpha_3} + g_{\alpha_2}g_{\alpha_3})
+ 4s\qty(\sqrt{2} - h_1(\alpha_1,\alpha_2,\alpha_3))
    \end{aligned}
\end{equation}
We need to minimize $\lambda_2(\alpha_1,\alpha_2,\alpha_3,s)$ and for that, $h_1(\alpha_1,\alpha_2,\alpha_3)$ has to be maximized. The maxima of $h_1(\alpha_1,\alpha_2,\alpha_3)$ is $\sqrt{2}$ attained at $\alpha_1=\alpha_2=\alpha_3=\frac{\pi}{4}$. Hence, the minimum value of the second eigenvalue is zero. Another argument for minimization is as follows $\lambda_{-}(\alpha_1,\alpha_2,\alpha_3,s)$ is symmetric under permutation of variables which means that optima will occur at $\alpha_1=\alpha_2=\alpha_3$ and
\begin{equation}
    \lambda_-(\alpha_1,\alpha_1,\alpha_1,s) = \frac{3}{4}(-1+g_{\alpha_1}^2) + 4s\qty[\sqrt{2}-(\cos(\alpha_1)+\sin(2\alpha_1)\sin(\alpha_1))]
\end{equation}
Thus the minima of this function occurs at the boundary of domain. The minimum value of the above is zero at $\alpha_1=\alpha_2=\alpha_3=0$ and $\alpha_1=\alpha_2=\alpha_3=\frac{\pi}{4}$ hence $Q_1\succeq0$.
{\subsubsection{Positivity of $Q_2$}
To show that $Q_2\succeq0$, we will show that both of it's eigenvalues are greater than or equal to zero. The two eigenvalues of $Q_2$ are simply $\lambda_{\pm}(\alpha_1,\alpha_2,\alpha_3,s)=f_3(\alpha_1,\alpha_2,\alpha_3,s)\pm f_4(\alpha_1,\alpha_2,\alpha_3,s)$. Considering the first eigenvalue,
\begin{equation}
    \begin{aligned}
        \lambda_+(\alpha_1,\alpha_2,\alpha_3,s) = f_3(\alpha_1,\alpha_2,\alpha_3,s)+ f_4(\alpha_1,\alpha_2,\alpha_3,s)
= -1 + 4s\qty(\sqrt{2} + \cos(\alpha_1-\alpha_2)\cos(\alpha_3) - \sin(\alpha_1+\alpha_2)\sin(\alpha_3))
    \end{aligned}
\end{equation}
The function $h_2(\alpha_1,\alpha_2,\alpha_3) = \cos(\alpha_1-\alpha_2)\cos(\alpha_3) - \sin(\alpha_1+\alpha_2)\sin(\alpha_3)$ is always non-negative. This can be proven by first showing that $\cos(\alpha_1-\alpha_2)\geq\sin(\alpha_1+\alpha_2)$.
\begin{equation}
\begin{aligned}
        \cos(\alpha_1-\alpha_2)-\sin(\alpha_1+\alpha_2)&=\cos(\alpha_1)\cos(\alpha_2)+\sin(\alpha_1)\sin(\alpha_2)-\sin(\alpha_1)\cos (a_2)-\cos(\alpha_1)\sin (a_2)\\
        &=\qty[\cos(\alpha_1)-\sin(\alpha_1)]\qty[\cos(\alpha_2)-\sin(\alpha_2)]
\end{aligned}
\end{equation}
since, in the interval $[0,\pi/4]$ $\cos(x)\geq\sin(x)$, $\qty[\cos(\alpha_1)-\sin(\alpha_1)]\qty[\cos(\alpha_2)-\sin(\alpha_2)]\geq0$. which implies that $h_2(\alpha_1,\alpha_2,\alpha_3)\geq0$. For minimising $\lambda_+(\alpha_1,\alpha_2,\alpha_3,s)$ we need minimum of $h_2(\alpha_1,\alpha_2,\alpha_3)$ which is zero. Hence, $\min(\lambda_+(\alpha_1,\alpha_2,\alpha_3,s)) = -1+4\sqrt{2}s$ which is $\frac{1}{4}\qty(2+3\sqrt{2})\geq0$ for $s=\frac{3}{16}\qty(1+\sqrt{2})$. Hence $\lambda_+(\alpha_1,\alpha_2,\alpha_3,s)\geq0$.
Now consider the second eigenvalue
\begin{equation}
    \begin{aligned}
        \lambda_-(\alpha_1,\alpha_2,\alpha_3,s) &= f_3(\alpha_1,\alpha_2,\alpha_3,s) - f_4(\alpha_1,\alpha_2,\alpha_3,s) \\
        &=\frac{1}{4}\qty(-3 + g_{\alpha_1}g_{\alpha_2} - g_{\alpha_1}g_{\alpha_3} - g_{\alpha_2}g_{\alpha_3})
+ 4s\qty(\sqrt{2} - h_2(\alpha_1,\alpha_2,\alpha_3))
    \end{aligned}
\end{equation}
The minimum of the above is zero which occurs at $\alpha_1=\alpha_2=\alpha_3=0$. Hence $Q_2\succeq0$.

\subsubsection{Positivity of $Q_3$}

To show that $Q_3\succeq0$, we will show that both of it's eigenvalues are greater than or equal to zero. The two eigenvalues of $Q_3$ are simply $\lambda_{\pm}(\alpha_1,\alpha_2,\alpha_3,s) = f_5(\alpha_1,\alpha_2,\alpha_3,s)\pm f_6(\alpha_1,\alpha_2,\alpha_3,s)$. Considering the first eigenvalue,
\begin{equation}
    \begin{aligned}
        \lambda_+(\alpha_1,\alpha_2,\alpha_3,s) &= f_5(\alpha_1,\alpha_2,\alpha_3,s)+ f_6(\alpha_1,\alpha_2,\alpha_3,s) = -1 + 4 s \qty(\sqrt{2} + \cos{(\alpha_1 + \alpha_2)} \cos{\alpha_3} + \sin{(\alpha_1 - \alpha_2)} \sin{\alpha_3})
    \end{aligned}
\end{equation}
The function $h_3(\alpha_1,\alpha_2,\alpha_3) = (\cos{(\alpha_1 + \alpha_2)} \cos{\alpha_3} + \sin{(\alpha_1 - \alpha_2)} \sin{\alpha_3})$ is minimized at the corners of the cube $\alpha_1,\alpha_2,\alpha_3\in[0,\frac{\pi}{4}]$ because in this domain, it's gradient is never zero inside the domain. Then, $\min(h_3(\alpha_1,\alpha_2,\alpha_3)) =  0$ which implies that $\min (\lambda_+) = -1+4\sqrt{2}s$ which is greater than zero for $s=\frac{3\qty(1+\sqrt{2})}{16}$. Now consider the second eigenvalue $\lambda_- = f_5(\alpha_1,\alpha_2,\alpha_3,s) - f_6(\alpha_1,\alpha_2,\alpha_3,s)$
\begin{equation}
\begin{aligned}
\lambda_-(\alpha_1,\alpha_2,\alpha_3,s) 
= \frac{1}{4} \Big(
-3 - g_{\alpha_2} g_{\alpha_3} 
+ g_{\alpha_1} (-g_{\alpha_2} + g_{\alpha_3}) 
+ 16 s \big( \sqrt{2} - h_3(\alpha_1,\alpha_2,\alpha_3) \big)
\Big)
\end{aligned}
\end{equation}
We need to minimize $\lambda_-(\alpha_1,\alpha_2,\alpha_3,s)$ and for that, $h_3(\alpha_1,\alpha_2,\alpha_3)$ has to be maximized. The maxima of $h_3(\alpha_1,\alpha_2,\alpha_3)$ is $1$ attained at $\alpha_1=\alpha_3=\frac{\pi}{4}, \alpha_2 = 0$. hence the minimum value of the second eigenvalue is zero, hence $Q_3\succeq0$.

\subsubsection{Positivity of $Q_4$}
To show that $Q_4\succeq0$, we will show that both of it's eigenvalues are greater than or equal to zero. The two eigenvalues of $Q_4$ are simply $\lambda_{\pm}(\alpha_1,\alpha_2,\alpha_3,s) = f_7(\alpha_1,\alpha_2,\alpha_3,s)\pm f_8(\alpha_1,\alpha_2,\alpha_3,s)$. Considering the first eigenvalue,
\begin{equation}
    \begin{aligned}
        \lambda_+(\alpha_1,\alpha_2,\alpha_3,s) &= f_7(\alpha_1,\alpha_2,\alpha_3,s)+ f_8(\alpha_1,\alpha_2,\alpha_3,s) = -1 + 4 s \qty(\sqrt{2} + \cos(\alpha_1 + \alpha_2) \cos(\alpha_3) - \sin(\alpha_1 - \alpha_2) \sin(\alpha_3))
    \end{aligned}
\end{equation}
The function $h_4(\alpha_1,\alpha_2,\alpha_3) = (\cos{(\alpha_1 + \alpha_2)} \cos{\alpha_3} - \sin{(\alpha_1 - \alpha_2)} \sin{\alpha_3})$ is minimized at the corners of the cube $\alpha_1,\alpha_2,\alpha_3\in[0,\frac{\pi}{4}]$ because in this domain, it's gradient is never zero inside the domain. Then, $\min(h_3(\alpha_1,\alpha_2,\alpha_3)) =  0$ which implies that $\min (\lambda_+) = -1+4\sqrt{2}s$ which is greater than zero for $s=\frac{3\qty(1+\sqrt{2})}{16}$. Now consider the second eigenvalue $\lambda_- = f_7(\alpha_1,\alpha_2,\alpha_3,s) - f_8(\alpha_1,\alpha_2,\alpha_3,s)$
\begin{equation}
\begin{aligned}
\lambda_-(\alpha_1,\alpha_2,\alpha_3,s) 
= \frac{1}{4}(-3 + g_{\alpha_2} g_{\alpha_3} - g_{\alpha_1} (g_{\alpha_2} + g_{\alpha_3}) + 
   16 s (\sqrt{2} - h_4(\alpha_1,\alpha_2,\alpha_3))
\end{aligned}
\end{equation}
We need to minimize $\lambda_-(\alpha_1,\alpha_2,\alpha_3,s)$ and for that, $h_4(\alpha_1,\alpha_2,\alpha_3)$ has to be maximized. The maxima of $h_4(\alpha_1,\alpha_2,\alpha_3)$ is $1$ attained at $ \alpha_1 = 0, \alpha_2=\alpha_3=\frac{\pi}{4}$, hence the minimum value of the second eigenvalue is zero, hence $Q_4\succeq0$.

\section{Robustness of the self-testing protocol based on 4-Partite Svetlichny inequality}\label{svet4}
For $n = 4$, by using Eq.~(\ref{eq:LocalMeasurements}) the Svetlichny functional in Eq.~(\ref{eq:ParameterisedSvOperator}) can be written as
\begin{equation}
\begin{aligned}
\mathcal{S}_4(\alpha_1,\alpha_2,\alpha_3,\alpha_4)
= 4\Big(&
-\cos(\alpha_1)\cos(\alpha_2)\cos(\alpha_3)\cos(\alpha_4)\,\sigma_x\!\otimes\!\sigma_x\!\otimes\!\sigma_x\!\otimes\!\sigma_x
+ \cos(\alpha_1)\cos(\alpha_2)\cos(\alpha_3)\sin(\alpha_4)\,\sigma_x\!\otimes\!\sigma_x\!\otimes\!\sigma_x\!\otimes\!\sigma_y
\\[-2pt]&\quad
+ \cos(\alpha_1)\cos(\alpha_2)\cos(\alpha_4)\sin(\alpha_3)\,\sigma_x\!\otimes\!\sigma_x\!\otimes\!\sigma_y\!\otimes\!\sigma_x
+ \cos(\alpha_1)\cos(\alpha_2)\sin(\alpha_3)\sin(\alpha_4)\,\sigma_x\!\otimes\!\sigma_x\!\otimes\!\sigma_y\!\otimes\!\sigma_y
\\[-2pt]&\quad
+ \cos(\alpha_1)\cos(\alpha_3)\cos(\alpha_4)\sin(\alpha_2)\,\sigma_x\!\otimes\!\sigma_y\!\otimes\!\sigma_x\!\otimes\!\sigma_x
+ \cos(\alpha_1)\cos(\alpha_3)\sin(\alpha_2)\sin(\alpha_4)\,\sigma_x\!\otimes\!\sigma_y\!\otimes\!\sigma_x\!\otimes\!\sigma_y
\\[2pt]&\quad
+ \cos(\alpha_1)\cos(\alpha_4)\sin(\alpha_2)\sin(\alpha_3)\,\sigma_x\!\otimes\!\sigma_y\!\otimes\!\sigma_y\!\otimes\!\sigma_x
- \cos(\alpha_1)\sin(\alpha_2)\sin(\alpha_3)\sin(\alpha_4)\,\sigma_x\!\otimes\!\sigma_y\!\otimes\!\sigma_y\!\otimes\!\sigma_y
\\[-2pt]&\quad
+ \cos(\alpha_2)\cos(\alpha_3)\cos(\alpha_4)\sin(\alpha_1)\,\sigma_y\!\otimes\!\sigma_x\!\otimes\!\sigma_x\!\otimes\!\sigma_x
+ \cos(\alpha_2)\cos(\alpha_3)\sin(\alpha_1)\sin(\alpha_4)\,\sigma_y\!\otimes\!\sigma_x\!\otimes\!\sigma_x\!\otimes\!\sigma_y
\\[-2pt]&\quad
+ \cos(\alpha_2)\cos(\alpha_4)\sin(\alpha_1)\sin(\alpha_3)\,\sigma_y\!\otimes\!\sigma_x\!\otimes\!\sigma_y\!\otimes\!\sigma_x
- \cos(\alpha_2)\sin(\alpha_1)\sin(\alpha_3)\sin(\alpha_4)\,\sigma_y\!\otimes\!\sigma_x\!\otimes\!\sigma_y\!\otimes\!\sigma_y
\\[2pt]&\quad
+ \cos(\alpha_3)\cos(\alpha_4)\sin(\alpha_1)\sin(\alpha_2)\,\sigma_y\!\otimes\!\sigma_y\!\otimes\!\sigma_x\!\otimes\!\sigma_x
- \cos(\alpha_3)\sin(\alpha_1)\sin(\alpha_2)\sin(\alpha_4)\,\sigma_y\!\otimes\!\sigma_y\!\otimes\!\sigma_x\!\otimes\!\sigma_y
\\[-2pt]&\quad
- \cos(\alpha_4)\sin(\alpha_1)\sin(\alpha_2)\sin(\alpha_3)\,\sigma_y\!\otimes\!\sigma_y\!\otimes\!\sigma_y\!\otimes\!\sigma_x
- \sin(\alpha_1)\sin(\alpha_2)\sin(\alpha_3)\sin(\alpha_4)\,\sigma_y\!\otimes\!\sigma_y\!\otimes\!\sigma_y\!\otimes\!\sigma_y
\Big)
\end{aligned}
\label{eq:W4Sv}
\end{equation}
The $4-$partite GHZ state which gives the maximum value $(\mathcal{S}_4)_Q = 8\sqrt{2}$ as in Eq.~(\ref{eq:RhoSn}) is given by
\begin{equation}
    \begin{aligned}
    \rho_{\mathcal{S}_4} =
    \; \frac{1}{16} \Big(
      &\sigma_z \otimes \sigma_z \otimes \sigma_z \otimes \sigma_z +
      \sigma_z \otimes \sigma_z \otimes \mathbb{I} \otimes \mathbb{I} +
      \sigma_z \otimes \mathbb{I} \otimes \sigma_z \otimes \mathbb{I} + \\
    &\quad \sigma_z \otimes \mathbb{I} \otimes \mathbb{I} \otimes \sigma_z +
      \mathbb{I} \otimes \sigma_z \otimes \sigma_z \otimes \mathbb{I} +
      \mathbb{I} \otimes \sigma_z \otimes \mathbb{I} \otimes \sigma_z +
      \mathbb{I} \otimes \mathbb{I} \otimes \sigma_z \otimes \sigma_z +
      \mathbb{I} \otimes \mathbb{I} \otimes \mathbb{I} \otimes \mathbb{I}
    \Big) \\
    &\; + \frac{1}{16\sqrt{2}} \Big(
      -\,\sigma_y \otimes \sigma_y \otimes \sigma_y \otimes \sigma_y
      -\,\sigma_y \otimes \sigma_y \otimes \sigma_y \otimes \sigma_x
      -\,\sigma_y \otimes \sigma_y \otimes \sigma_x \otimes \sigma_y \\
    &\quad +\,\sigma_y \otimes \sigma_y \otimes \sigma_x \otimes \sigma_x
      -\,\sigma_y \otimes \sigma_x \otimes \sigma_y \otimes \sigma_y
      +\,\sigma_y \otimes \sigma_x \otimes \sigma_y \otimes \sigma_x
      +\,\sigma_y \otimes \sigma_x \otimes \sigma_x \otimes \sigma_y \\
    &\quad +\,\sigma_y \otimes \sigma_x \otimes \sigma_x \otimes \sigma_x
      -\,\sigma_x \otimes \sigma_y \otimes \sigma_y \otimes \sigma_y
      +\,\sigma_x \otimes \sigma_y \otimes \sigma_y \otimes \sigma_x
      +\,\sigma_x \otimes \sigma_y \otimes \sigma_x \otimes \sigma_y \\
    &\quad +\,\sigma_x \otimes \sigma_y \otimes \sigma_x \otimes \sigma_x
      +\,\sigma_x \otimes \sigma_x \otimes \sigma_y \otimes \sigma_y
      +\,\sigma_x \otimes \sigma_x \otimes \sigma_y \otimes \sigma_x
      +\,\sigma_x \otimes \sigma_x \otimes \sigma_x \otimes \sigma_y \\
    &\quad -\,\sigma_x \otimes \sigma_x \otimes \sigma_x \otimes \sigma_x
    \Big)
\end{aligned}
\label{4svs}
\end{equation}
After applying the dephasing channel defined in Eq.~(\ref{eq:DephasingChannel}), the state (\ref{4svs}) becomes
\begin{equation}
\begin{aligned}
K(\alpha_1,\alpha_2,\alpha_3,\alpha_4) &= \frac{1}{16} \Big( \;
g_{\alpha_1}g_{\alpha_2}g_{\alpha_3}g_{\alpha_4}\, \sigma_z \otimes \sigma_z \otimes \sigma_z \otimes \sigma_z
+ g_{\alpha_1}g_{\alpha_2}\, \sigma_z \otimes \sigma_z \otimes \mathbb{I} \otimes \mathbb{I} \\[4pt]
&\quad
+ g_{\alpha_1}g_{\alpha_3}\, \sigma_z \otimes \mathbb{I} \otimes \sigma_z \otimes \mathbb{I}
+ g_{\alpha_1}g_{\alpha_4}\, \sigma_z \otimes \mathbb{I} \otimes \mathbb{I} \otimes \sigma_z
+ g_{\alpha_2}g_{\alpha_3}\, \mathbb{I} \otimes \sigma_z \otimes \sigma_z \otimes \mathbb{I} \\[4pt]
&\quad
+ g_{\alpha_2}g_{\alpha_4}\, \mathbb{I} \otimes \sigma_z \otimes \mathbb{I} \otimes \sigma_z
+ g_{\alpha_3}g_{\alpha_4}\, \mathbb{I} \otimes \mathbb{I} \otimes \sigma_z \otimes \sigma_z
+ \mathbb{I} \otimes \mathbb{I} \otimes \mathbb{I} \otimes \mathbb{I}
\;\Big) \\[6pt]
&\quad + \frac{1}{16\sqrt{2}} \Big( \;
- g_{\alpha_1}g_{\alpha_2}g_{\alpha_3}g_{\alpha_4}\, \sigma_y \otimes \sigma_y \otimes \sigma_y \otimes \sigma_y
- g_{\alpha_1}g_{\alpha_2}g_{\alpha_3}\, \sigma_y \otimes \sigma_y \otimes \sigma_y \otimes \sigma_x \\[4pt]
&\qquad
- g_{\alpha_1}g_{\alpha_2}g_{\alpha_4}\, \sigma_y \otimes \sigma_y \otimes \sigma_x \otimes \sigma_y
+ g_{\alpha_1}g_{\alpha_2}\, \sigma_y \otimes \sigma_y \otimes \sigma_x \otimes \sigma_x
- g_{\alpha_1}g_{\alpha_3}g_{\alpha_4}\, \sigma_y \otimes \sigma_x \otimes \sigma_y \otimes \sigma_y \\[4pt]
&\qquad
+ g_{\alpha_1}g_{\alpha_3}\, \sigma_y \otimes \sigma_x \otimes \sigma_y \otimes \sigma_x
+ g_{\alpha_1}g_{\alpha_4}\, \sigma_y \otimes \sigma_x \otimes \sigma_x \otimes \sigma_y
+ g_{\alpha_1}\, \sigma_y \otimes \sigma_x \otimes \sigma_x \otimes \sigma_x \\[4pt]
&\qquad
- g_{\alpha_2}g_{\alpha_3}g_{\alpha_4}\, \sigma_x \otimes \sigma_y \otimes \sigma_y \otimes \sigma_y
+ g_{\alpha_2}g_{\alpha_3}\, \sigma_x \otimes \sigma_y \otimes \sigma_y \otimes \sigma_x
+ g_{\alpha_2}g_{\alpha_4}\, \sigma_x \otimes \sigma_y \otimes \sigma_x \otimes \sigma_y \\[4pt]
&\qquad
+ g_{\alpha_2}\, \sigma_x \otimes \sigma_y \otimes \sigma_x \otimes \sigma_x
+ g_{\alpha_3}g_{\alpha_4}\, \sigma_x \otimes \sigma_x \otimes \sigma_y \otimes \sigma_y
+ g_{\alpha_3}\, \sigma_x \otimes \sigma_x \otimes \sigma_y \otimes \sigma_x \\[4pt]
&\qquad
+ g_{\alpha_4}\, \sigma_x \otimes \sigma_x \otimes \sigma_x \otimes \sigma_y
- \sigma_x \otimes \sigma_x \otimes \sigma_x \otimes \sigma_x
\;\Big)
\end{aligned}
\label{eq:K4Sv}
\end{equation}
To prove the robustness claim, we have to prove $T_4=\mathcal{K}_{\mathcal{S}_4}(\alpha_1,\alpha_2,\alpha_3,\alpha_4)-s\mathcal{S}_4(\alpha_1,\alpha_2,\alpha_3,\alpha_4)-\mu\mathbb{I}\otimes\mathbb{I}\otimes\mathbb{I}\otimes\mathbb{I}\succeq0$. Now, the operator, $T_4 $ is a sparse  persymmetric matrix which has non-zero diagonal and antidiagonal entries. We consider a unitary matrix of the following form
Consider the unitary matrix of the following form
\begin{equation}
    U = \sum_{k=0}^7 \ket{2k}\bra{k} +\sum_{k=8}^{15} \ket{31-2k}\bra{k}
\end{equation}
where, $\ket{i}$ is the standard Dirac notation. Now, $UT_4U^\dag$ is a block diagonal matrix with eight $2\times2$ blocks along its diagonals.
\begin{equation}
    UT_4U^\dag = Q_1 \oplus Q_2\oplus Q_3 \oplus Q_4 \oplus Q_5 \oplus Q_6 \oplus Q_7 \oplus Q_8
\end{equation}
Now, proving positivity of $T_4$ requires proving positivity of each of the $Q_i$. Consider
\begin{equation}
   Q_1 = \begin{bmatrix}
        f_1(\alpha_1,\alpha_2,\alpha_3,\alpha_4,s) & f_2(\alpha_1,\alpha_2,\alpha_3,\alpha_4,s)\\
        f_2(\alpha_1,\alpha_2,\alpha_3,\alpha_4,s)^* & f_1(\alpha_1,\alpha_2,\alpha_3,\alpha_4,s)
    \end{bmatrix}
\end{equation}
where, $f_2(\alpha_1,\alpha_2,\alpha_3,\alpha_4,s)^*$ is complex conjugate of $f_2(\alpha_1,\alpha_2,\alpha_3,\alpha_4,s)$. Also, 
\begin{equation}
    \begin{aligned}
       f_1(\alpha_1,\alpha_2,\alpha_3,\alpha_4,s) &= \frac{1}{16}\qty(
-15 + g_{\alpha_1}g_{\alpha_2} + g_{\alpha_1}g_{\alpha_3} + g_{\alpha_2}g_{\alpha_3}
+ g_{\alpha_1}g_{\alpha_4} + g_{\alpha_2}g_{\alpha_4} + g_{\alpha_3}g_{\alpha_4}
+ g_{\alpha_1}g_{\alpha_2}g_{\alpha_3}g_{\alpha_4}
) + 8\sqrt{2}s \\[6pt]
f_2(\alpha_1,\alpha_2,\alpha_3,\alpha_4,s) &= \frac{1}{16\sqrt{2}}\qty(
-1 - g_{\alpha_1}g_{\alpha_2} - g_{\alpha_1}g_{\alpha_3} - g_{\alpha_2}g_{\alpha_3}
- g_{\alpha_1}g_{\alpha_4} - g_{\alpha_2}g_{\alpha_4} - g_{\alpha_3}g_{\alpha_4}
- g_{\alpha_1}g_{\alpha_2}g_{\alpha_3}g_{\alpha_4}
) \\
&\quad + 4s \qty[
\cos(\alpha_1-\alpha_4)\cos(\alpha_2-\alpha_3)
+ \sin(\alpha_2+\alpha_3)\sin(\alpha_1+\alpha_4)
] \\
&\quad + i\bigg(
\frac{1}{16\sqrt{2}}\qty(
g_{\alpha_1}+g_{\alpha_2}+g_{\alpha_3}+g_{\alpha_4}
+ g_{\alpha_1}g_{\alpha_2}g_{\alpha_3}
+ g_{\alpha_1}g_{\alpha_2}g_{\alpha_4}
+ g_{\alpha_1}g_{\alpha_3}g_{\alpha_4}
+ g_{\alpha_2}g_{\alpha_3}g_{\alpha_4}
) \\
&\qquad - 4s \qty[
\cos(\alpha_1-\alpha_4)\sin(\alpha_2+\alpha_3)
+ \cos(\alpha_2-\alpha_3)\sin(\alpha_1+\alpha_4)
]
\bigg).
    \end{aligned}
\end{equation}
We now demonstrate that $Q_1$ is positive by applying Sylvester's criterion \cite{Boyd2004}. Using this criteria, to prove that $Q_1$ is positive semi-definite, we need to show that $f_1\qty(\alpha_1,\alpha_2,\alpha_3,\alpha_4,s)\geq0$ and $\qty(f_1\qty(\alpha_1,\alpha_2,\alpha_3,\alpha_4,s))^2 - \abs{f_2\qty(\alpha_1,\alpha_2,\alpha_3,\alpha_4,s)}^2\geq0$. Firstly, we have to prove that $f_1(\alpha_1,\alpha_2,\alpha_3,\alpha_4,s)$ is always positive for $s\geq\frac{1+\sqrt{2}}{16}$. To prove this, consider the following optimization problem 
\begin{align}
\text{minimize} \quad & f_1(\alpha_1,\alpha_2,\alpha_3,\alpha_4,s) \\
\text{subject to} \quad & 0\leq \alpha_1,\alpha_2,\alpha_3,\alpha_4\leq \frac{\pi}{4} \\
& s = \frac{1+\sqrt{2}}{16}
\end{align}
Since $g(x)$ is a bijective function in the range $[0,\frac{\pi}{4}]$ a change of variable by replacing $g_{\alpha_1} \to x_1,g_{\alpha_2} \to x_2,g_{\alpha_3} \to x_3,g_{\alpha_4} \to x_4$ gives the following equivalent optimization problem \cite{Boyd2004}
\begin{align*}
\text{minimize} \quad &  \frac{1}{16}\qty(-15+x_1x_2+x_1x_3+x_2x_3+x_1x_4++x_2x_4+x_3x_4+x_1x_2x_3x_4)+8\sqrt{2}s \\
\text{subject to} \quad & 0\leq x_i\leq 1 \quad i\in\{1,2,3,4\}\\
& s = \frac{1+\sqrt{2}}{16}
\end{align*}
The above problem is a multilinear optimization problem over a cube hence the minima occurs at one of it's corners (\textit{Strong Maximum Principle for Harmonic functions} \cite{Axler2001}). The minima occurs at $x_i=0$. Hence $\min\big(f_1(\alpha_1,\alpha_2,\alpha_3,\alpha_4,s)\big)=\frac{1+8\sqrt{2}}{16}$. Thus, $f_1(\alpha_1,\alpha_2,\alpha_3,\alpha_4,s)$ is always positive. Now, we have
\begin{equation}
    \begin{aligned}
        f_1&= \frac{1}{16}(-15+g_e)+8\sqrt{2}s\\
        f_2&= -\frac{1}{16\sqrt{2}}(1+g_e) + 4t_1s + i\qty(\frac{g_o}{16\sqrt{2}}+4t_2s)
    \end{aligned}
\end{equation}
where
\begin{equation}
\begin{aligned}
        g_e &= g_{\alpha_1}g_{\alpha_2} + g_{\alpha_1}g_{\alpha_3} + g_{\alpha_2}g_{\alpha_3}
+ g_{\alpha_1}g_{\alpha_4} + g_{\alpha_2}g_{\alpha_4} + g_{\alpha_3}g_{\alpha_4}
+ g_{\alpha_1}g_{\alpha_2}g_{\alpha_3}g_{\alpha_4} \\[6pt]
g_o &= g_{\alpha_1} + g_{\alpha_2} + g_{\alpha_3} + g_{\alpha_4}
+ g_{\alpha_1}g_{\alpha_2}g_{\alpha_3}
+ g_{\alpha_1}g_{\alpha_2}g_{\alpha_4}
+ g_{\alpha_1}g_{\alpha_3}g_{\alpha_4}
+ g_{\alpha_2}g_{\alpha_3}g_{\alpha_4} \\
        t_1 &= \cos(\alpha_1-\alpha_4)\cos(\alpha_2-\alpha_3) + \sin(\alpha_2+\alpha_3)\sin(\alpha_1+\alpha_4)\\
        t_2 &= -\cos(\alpha_1-\alpha_4)\sin(\alpha_2+\alpha_3) - \cos(\alpha_2-\alpha_3)\sin(\alpha_1+\alpha_4)
\end{aligned}
\end{equation}
Now, we evaluate 
\begin{equation}\label{deter}
\begin{aligned}
        f_1^2-\abs{f_2}^2 &= \qty[128-16(t_1^2+t_2^2)]s^2+\qty[-15\sqrt{2}+g_e\sqrt{2}+\frac{1+g_e}{2\sqrt{2}}t_1-\frac{g_o}{2\sqrt{2}}t_2]s \\
        &+ \frac{1}{16^2}\qty[\frac{(g_e+g_o)(g_e-g_o)}{2}-31g_e+\frac{449}{2}]
\end{aligned}
\end{equation}
The minimum of the above function is zero at $\alpha_1=\alpha_2=\alpha_3 = \alpha_4$. Thus $Q_1\succeq0$.
 The positivity of the other seven blocks can also be proven similarly to establish that $T_4$ is positive semi-definite, which establishes the theorem for the four partite Svetlichny inequality. 
\section{Alternative proof for Robustness of the protocol based on 3-Partite Svetlichny inequality}
Our goal is to find $s$ and $\mu$ such that Eq. (\ref{eq: Sv Operator Inequality}) holds for $n=3$, or equivalently
\begin{equation}
\begin{aligned}
 T_3(\alpha_1,\alpha_2,\alpha_3) &= \mathcal{K}_{\mathcal{S}_3}(\alpha_1,\alpha_2,\alpha_3) - s\mathcal{S}_3(\alpha_1,\alpha_2,\alpha_3) - \mu \mathbb{I} \\[6pt]
&= \left( \frac{1}{8} - \mu \right)\, \mathbb{I} \otimes \mathbb{I} \otimes \mathbb{I}
    + \frac{1}{8} \Big[
        g_{\alpha_2}g_{\alpha_3}\, \mathbb{I} \otimes \sigma_z \otimes \sigma_z
        + g_{\alpha_1}g_{\alpha_3}\, \sigma_z \otimes \mathbb{I} \otimes \sigma_z
        + g_{\alpha_1}g_{\alpha_2}\, \sigma_z \otimes \sigma_z \otimes \mathbb{I}
    \Big] \\[6pt]
&\quad + \mathcal{T}_1(\alpha_1,\alpha_2,\alpha_3)\, \sigma_x \otimes \sigma_x \otimes \sigma_x
    + \mathcal{T}_2(\alpha_1,\alpha_2,\alpha_3)\, \sigma_x \otimes \sigma_y \otimes \sigma_y \\[6pt]
&\quad + \mathcal{T}_3(\alpha_1,\alpha_2,\alpha_3)\, \sigma_y \otimes \sigma_x \otimes \sigma_y
    + \mathcal{T}_4(\alpha_1,\alpha_2,\alpha_3)\, \sigma_y \otimes \sigma_y \otimes \sigma_x \succeq 0
\end{aligned}
\label{eq:T3}
\end{equation}
}
where
\begin{equation}
    \begin{aligned}
       &\mathcal{T}_1(\alpha_1,\alpha_2,\alpha_3)
        = -\frac{1}{8}
        + 4s\,\cos(\alpha_1)\cos(\alpha_2)\cos(\alpha_3),
        \qquad
        &\mathcal{T}_2(\alpha_1,\alpha_2,\alpha_3)
        = \frac{1}{8}\,g_{\alpha_2}\,g_{\alpha_3}
        - 4s\,\cos(\alpha_1)\sin(\alpha_2)\sin(\alpha_3),
        \\[6pt]
        &\mathcal{T}_3(\alpha_1,\alpha_2,\alpha_3)
        = \frac{1}{8}\,g_{\alpha_1}\,g_{\alpha_3}
        - 4s\,\sin(\alpha_1)\cos(\alpha_2)\sin(\alpha_3),
        \qquad
        &\mathcal{T}_4(\alpha_1,\alpha_2,\alpha_3)
        = \frac{1}{8}\,g_{\alpha_1}\,g_{\alpha_2}
        - 4s\,\sin(\alpha_1)\sin(\alpha_2)\cos(\alpha_3)
    \end{aligned}
    \label{eq:T1,T2,T3,T4}
\end{equation}
Noticing that $[T_3(\alpha_1,\alpha_2,\alpha_3),\sigma_z\otimes\sigma_z\otimes\mathbb{I}]
=
[T_3(\alpha_1,\alpha_2,\alpha_3),\sigma_z\otimes\mathbb{I}\otimes\sigma_z]
=
[T_3(\alpha_1,\alpha_2,\alpha_3),\mathbb{I}\otimes\sigma_z\otimes\sigma_z]
= 0$ leads us to consider the projector
\begin{equation}
P_{x_1 x_2}=
\frac{1}{4}\Big(
\mathbb{I} \otimes \mathbb{I} \otimes \mathbb{I}
\;+\;
(-1)^{x_1}\, \sigma_z \otimes \sigma_z \otimes \mathbb{I}
\;+\;
(-1)^{x_2}\, \sigma_z \otimes \mathbb{I} \otimes \sigma_z
\;+\;
(-1)^{x_1+x_2}\, \mathbb{I} \otimes \sigma_z \otimes \sigma_z
\Big)
\end{equation}
for $x_1,x_2 \in \{0,1\}$. Now, in \cite{Kaniewski2016}, it was shown that the inequality in Eq.~(\ref{eq:T3}) holds iff 
\begin{equation}
    \lambda_{x_1 x_2}(\alpha_1,\alpha_2,\alpha_3) = \Tr{M_{x_1x_2}^2} - \left(\Tr{M_{x_1x_2}}\right)^2 \geq 0
\end{equation}
where $ M_{x_1x_2} = P_{x_1x_2}T_3(\alpha_1,\alpha_2,\alpha_3)$ then,
\begin{equation}
    \begin{aligned}
        \mathrm{Tr}\{M_{x_1x_2}\} &= \mathrm{Tr}[P_{x_1x_2}T_3(\alpha_1,\alpha_2,\alpha_3)] \\
        &= 2\bigg(\frac{1}{8} - \mu\bigg)
        + \frac{1}{4}\big[(-1)^{x_1}g_{\alpha_1}g_{\alpha_2}
        + (-1)^{x_2}g_{\alpha_1}g_{\alpha_3}
        + (-1)^{x_1+x_2}g_{\alpha_2}g_{\alpha_3}\big]
    \end{aligned}
\end{equation}
and 
\begin{equation}
\begin{aligned}
\operatorname{Tr}\{M^2_{x_1 x_2}\}
&=
2\left(\frac{1}{8}-\mu\right)^2
+ 2\left( \mathcal{T}_1^2 + \mathcal{T}_2^2 + \mathcal{T}_3^2 + \mathcal{T}_4^2 \right)
+ \frac{1}{32}\Big(
g_{\alpha_1}^2 g_{\alpha_2}^2 + g_{\alpha_2}^2 g_{\alpha_3}^2 + g_{\alpha_3}^2 g_{\alpha_1}^2
\Big)
\\[10pt]
&\quad
+ \frac{(-1)^{x_1}}{2}\Bigg[
\left(\left(\frac{1}{8} - \mu \right) + \frac{1}{8}g_{\alpha_3}^2\right) g_{\alpha_1} g_{\alpha_2}
+ 4\big(\mathcal{T}_2 \mathcal{T}_3 - \mathcal{T}_1 \mathcal{T}_4\big)
\Bigg]
\\[10pt]
&\quad
+ \frac{(-1)^{x_2}}{2}\Bigg[
\left(\left(\frac{1}{8} - \mu \right) + \frac{1}{8}g_{\alpha_2}^2\right) g_{\alpha_1} g_{\alpha_3}
+ 4\big(\mathcal{T}_2 \mathcal{T}_4 - \mathcal{T}_1 \mathcal{T}_3\big)
\Bigg]
\\[10pt]
&\quad
+ \frac{(-1)^{x_1+x_2}}{2}\Bigg[
\left(\left(\frac{1}{8} - \mu \right) + \frac{1}{8}g_{\alpha_1}^2\right) g_{\alpha_2} g_{\alpha_3}
+ 4\big(\mathcal{T}_3 \mathcal{T}_4 - \mathcal{T}_1 \mathcal{T}_2\big)
\Bigg]
\end{aligned}
\end{equation}
So our problem is to find the value of $s$ and $\mu$ such that
\begin{equation}
\begin{aligned}
\lambda_{x_1 x_2}(\alpha_1,\alpha_2,\alpha_3) &=2\left(\frac{1}{8}-\mu\right)^2
- 2\left( \mathcal{T}_1^2 + \mathcal{T}_2^2 + \mathcal{T}_3^2 + \mathcal{T}_4^2 \right)
+ \frac{1}{32}\Big(
g_{\alpha_1}^2 g_{\alpha_2}^2 + g_{\alpha_2}^2 g_{\alpha_3}^2 + g_{\alpha_3}^2 g_{\alpha_1}^2
\Big)
\\[10pt]
&\quad
+ (-1)^{x_1}\Bigg[
\frac{1}{2}\left(\left(\frac{1}{8} - \mu \right) + \frac{1}{8}g_{\alpha_3}^2\right) g_{\alpha_1} g_{\alpha_2}
- 4\big(\mathcal{T}_2 \mathcal{T}_3 - \mathcal{T}_1 \mathcal{T}_4\big)
\Bigg]
\\[10pt]
&\quad
+ (-1)^{x_2}\Bigg[
\frac{1}{2}\left(\left(\frac{1}{8} - \mu \right) + \frac{1}{8}g_{\alpha_2}^2\right) g_{\alpha_1} g_{\alpha_3}
- 4\big(\mathcal{T}_2 \mathcal{T}_4 - \mathcal{T}_1 \mathcal{T}_3\big)
\Bigg]
\\[10pt]
&\quad
+ (-1)^{x_1+x_2}\Bigg[
\frac{1}{2}\left(\left(\frac{1}{8} - \mu \right) + \frac{1}{8}g_{\alpha_1}^2\right) g_{\alpha_2} g_{\alpha_3}
- 4\big(\mathcal{T}_3 \mathcal{T}_4 - \mathcal{T}_1 \mathcal{T}_2\big)
\Bigg] \ \ \geq 0
\end{aligned}
\label{eq:lambda}
\end{equation}
holds $\forall (\alpha_1,\alpha_2,\alpha_3) \in [0,\pi/4], \; x_1,x_2 \in \{0,1\}$.
{\subsubsection{Solving for the constants $s$ and $\mu$}
From the self-testing statement in Eq.~(\ref{eq:f_beta}), we have
\begin{equation}
    \mu = 1 - 4\sqrt{2}s
\label{eq:s}
\end{equation}
Since the fidelity of the unknown state giving the maximum quantum value should be $1$.
Consider the case $x_1=0, x_2 = 1$, our purpose is to prove that Eq. (\ref{eq:lambda})
\begin{equation}
\begin{aligned}
    \lambda_{01}(\alpha_1,\alpha_2,\alpha_3,s)
=
\frac{1}{2}
&\Bigl[
-1 + 4s\bigl(\sqrt{2} + X(\alpha_1,\alpha_2,\alpha_3) - Y(\alpha_1,\alpha_2,\alpha_3)\bigr)
\Bigr] \\
&\Bigl[
-3 + g_{\alpha_1}g_{\alpha_2} - \bigl(g_{\alpha_1}+g_{\alpha_2}\bigr)g_{\alpha_3}
+ 16s\bigl(\sqrt{2} - X(\alpha_1,\alpha_2,\alpha_3) + Y(\alpha_1,\alpha_2,\alpha_3)\bigr)
\Bigr]
\geq 0 ,
\end{aligned}
\label{eq:lambda00}
\end{equation}
Here we have used Eq.~\eqref{eq:s}, and
\begin{equation}
\begin{aligned}
X(\alpha_1,\alpha_2,\alpha_3) = \cos(\alpha_1-\alpha_2)\cos(\alpha_3), \qquad
Y(\alpha_1,\alpha_2,\alpha_3) = \sin(\alpha_1+\alpha_2)\sin(\alpha_3).
\end{aligned}
\label{eq:shorthandlambda00}
\end{equation}
For any $(\alpha_1,\alpha_2,\alpha_3)$, the expression $\lambda_{01}(\alpha_1,\alpha_2,\alpha_3,s)$ is a quadratic polynomial in $s$, with leading coefficient
\begin{equation}
c_{s^2}
= \operatorname{coeff}_{s^2}\!\left[\lambda_{01}\right]
= 32\bigl(2 - f^2(\alpha_1,\alpha_2,\alpha_3)\bigr),
\label{eq:coeffs^2}
\end{equation}
where we define $f(\alpha_1,\alpha_2,\alpha_3) = X(\alpha_1,\alpha_2,\alpha_3) - Y(\alpha_1,\alpha_2,\alpha_3) .$ Thus, for any $(\alpha_1,\alpha_2,\alpha_3)$, $\lambda_{01}(\alpha_1,\alpha_2,\alpha_3,s)$ represents either an upward- or downward-opening parabola in $s$, depending on the sign of $c_{s^2}$, which in turn is determined by $\max f^2(\alpha_1,\alpha_2,\alpha_3)$.
To determine the extrema of $f(\alpha_1,\alpha_2,\alpha_3)$ in the interior of the domain $\alpha_1, \alpha_2, \alpha_3 \in (0,\pi/4)$, we impose the stationary-point condition $\nabla f(\alpha_1,\alpha_2,\alpha_3) = 0$, i.e.,
\begin{equation}
\begin{aligned}
\partial_{\alpha_1} f(\alpha_1,\alpha_2,\alpha_3)
&= -\sin(\alpha_1-\alpha_2)\cos(\alpha_3) - \cos(\alpha_1+\alpha_2)\sin(\alpha_3) = 0, \\[6pt]
\partial_{\alpha_2} f(\alpha_1,\alpha_2,\alpha_3)
&= \sin(\alpha_1-\alpha_2)\cos(\alpha_3) - \cos(\alpha_1+\alpha_2)\sin(\alpha_3) = 0, \\[6pt]
\partial_{\alpha_3} f(\alpha_1,\alpha_2,\alpha_3)
&= -\cos(\alpha_1-\alpha_2)\sin(\alpha_3) - \sin(\alpha_1+\alpha_2)\cos(\alpha_3) = 0.
\end{aligned}
\label{eq:gradf}
\end{equation}
Adding the first two equations in Eq.~\eqref{eq:gradf}, we obtain
\begin{equation}
\cos(\alpha_1+\alpha_2)\sin(\alpha_3) = 0 ,
\end{equation}
which has no solution in the interior of the domain. Hence, $f(\alpha_1,\alpha_2,\alpha_3)$ does not admit any interior extremum, and its extrema must lie on the boundary. A direct analysis of the boundary shows that $\max f(\alpha_1,\alpha_2,\alpha_3) = 1$, attained at $\alpha_1=\alpha_2=\alpha_3 = 0$ and at $\alpha_1=\alpha_2=\frac{\pi}{4}, \alpha_3=0$. Consequently, $f^2(\alpha_1,\alpha_2,\alpha_3) \leq 1$ throughout the domain, and therefore
\begin{equation}
\min c_{s^2} = 32 > 0 .
\end{equation}
It follows that, for all admissible $(\alpha_1, \alpha_2, \alpha_3)$, $\lambda_{01}(\alpha_1, \alpha_2, \alpha_3,s)$ is an upward-opening parabola in $s$. Hence, the inequality \eqref{eq:lambda00} is satisfied for
\begin{equation}
s \in
\bigl(0,\min\{\min s_1,\min s_2\}\bigr]
\;\cup\;
\bigl[\max\{\max s_1,\max s_2\},\infty\bigr),
\end{equation}
where $s_1$ and $s_2$ denote the roots of $\lambda_{01}(\alpha_1, \alpha_2, \alpha_3,s)$. Solving for the roots using, we obtain
\begin{equation}
\begin{aligned}
s_1(\alpha_1,\alpha_2,\alpha_3)
&= \frac{1}{4\bigl(\sqrt{2} + X(\alpha_1,\alpha_2,\alpha_3) - Y(\alpha_1,\alpha_2,\alpha_3)\bigr)}, \\[6pt]
s_2(\alpha_1,\alpha_2,\alpha_3)
&= \frac{3 - g_{\alpha_1}g_{\alpha_2} + \bigl(g_{\alpha_1}+g_{\alpha_2}\bigr)g_{\alpha_3}}
{16\bigl(\sqrt{2} - X(\alpha_1,\alpha_2,\alpha_3) + Y(\alpha_1,\alpha_2,\alpha_3)\bigr)}
\end{aligned}
\label{eq:rootslambda}
\end{equation}
It is straightforward to see that
\begin{equation}
\begin{aligned}
\max s_1 &= \frac{1}{4\sqrt{2}},
\quad \text{attained at } \alpha_1=\alpha_2,\; \alpha_3=0, \\[6pt]
\min s_1 &= \frac{1}{4(\sqrt{2}+1)},
\quad \text{attained at } \alpha_1=\alpha_2=\alpha_3=\frac{\pi}{4}.
\end{aligned}
\end{equation}
Now we will prove that there are no critical points of $s_2$ in the interior of the domain $\alpha_1,\alpha_2,\alpha_3 \in (0,\pi/4)$.
\subsubsection*{Analysis of Critical Points of $s_2(\alpha_1,\alpha_2,\alpha_3)$}
Let us write $s_2(\alpha_1,\alpha_2,\alpha_3)$ as
\begin{equation}
    s_2(\alpha_1,\alpha_2,\alpha_3) = \frac{N(\alpha_1,\alpha_2,\alpha_3)}{D(\alpha_1,\alpha_2,\alpha_3)}
\end{equation}
where
\begin{equation}
\begin{aligned}
     N(\alpha_1,\alpha_2,\alpha_3) &= 3 - g_{\alpha_1}g_{\alpha_2} + (g_{\alpha_1} + g_{\alpha_2})g_{\alpha_3} \\
     D(\alpha_1,\alpha_2,\alpha_3) &=16\left(\sqrt{2}- \cos(\alpha_1-\alpha_2)\cos(\alpha_3)+ \sin(\alpha_1+\alpha_2)\sin(\alpha_3)\right)
\end{aligned}
\end{equation}
Since $D>0$, any critical point satisfies
\begin{equation}
\nabla s_2(\alpha_1,\alpha_2,\alpha_3) = 0
\quad \Longleftrightarrow \quad
(\nabla N)D - (\nabla D)N = 0
\label{eq:criticality}
\end{equation}
We compute
\begin{equation}
    \begin{aligned}
        \partial_{\alpha_1} N &= g'_{\alpha_3}\bigl(g_{\alpha_3}-g_{\alpha_2}\bigr), & \partial_{\alpha_1} D &= 16\bigl(\sin(\alpha_1-\alpha_2)\cos(\alpha_3) + \cos(\alpha_1+\alpha_2)\sin(\alpha_3)\bigr), \\
\partial_{\alpha_2} N &= g'_{\alpha_3}\bigl(g_{\alpha_3}-g_{\alpha_1}\bigr), & \partial_{\alpha_2} D &= 16\bigl(-\sin(\alpha_1-\alpha_2)\cos(\alpha_3) + \cos(\alpha_1+\alpha_2)\sin(\alpha_3)\bigr), \\
\partial_{\alpha_3} N &= g'_{\alpha_3}\bigl(g_{\alpha_1}+g_{\alpha_2}\bigr), & \partial_{\alpha_3} D &= 16\bigl(\cos(\alpha_1-\alpha_2)\sin(\alpha_3) + \sin(\alpha_1+\alpha_2)\cos(\alpha_3)\bigr).
    \end{aligned}
\end{equation}
where $\partial_{\alpha_i}N = \frac{\partial N}{\partial\alpha_i}$ and $\partial_{\alpha_i}D = \frac{\partial D}{\partial\alpha_i}$.
\subsection*{Eliminating $\alpha_1 \neq \alpha_2$}
The $\alpha_1$- and $\alpha_2$- components of Eq. (\ref{eq:criticality}) are,
\begin{align}
(\partial_{\alpha_1} N)D - (\partial_{\alpha_1} D)N = 0,
\qquad
(\partial_{\alpha_2} N)D - (\partial_{\alpha_2} D)N = 0\label{eq2}
\end{align}
Subtracting $\alpha_2-$Eq. from $\alpha_1-$Eq., we obtain
\begin{equation}
    \bigl[g'_{\alpha_3}\bigl(g_{\alpha_3}-g_{\alpha_2}\bigr) - g'_{\alpha_3}\bigl(g_{\alpha_3}-g_{\alpha_1}\bigr)\bigr]D
- 32N \sin(\alpha_1-\alpha_2)\cos(\alpha_3) = 0
\label{eq:D28}
\end{equation}
Define
\begin{equation}
    \mathcal{F}(\alpha_1,\alpha_2,\alpha_3)
= \bigl[g'_{\alpha_1}\bigl(g_{\alpha_3}-g_{\alpha_2}\bigr) - g'_{\alpha_2}\bigl(g_{\alpha_3}-g_{\alpha_1}\bigr)\bigr]D
- 32N \sin(\alpha_1-\alpha_2)\cos(\alpha_3)
\label{eq:mathcalF}
\end{equation}
Hence, Eq. (\ref{eq:D28}) holds when $\mathcal{F}(\alpha_1,\alpha_2,\alpha_3) = 0$. It is clear that for $\alpha_1=\alpha_2$, $\mathcal{F}(\alpha_1,\alpha_2,\alpha_3)=0$, hence $(\alpha_1-\alpha_2)$ is a factor of $\mathcal{F}(\alpha_1,\alpha_2,\alpha_3)$. So, we can write
\begin{equation}
    \mathcal{F}(\alpha_1,\alpha_2,\alpha_3) = (\alpha_1-\alpha_2)\,\Phi(\alpha_1,\alpha_2,\alpha_3)
\end{equation}
where
\begin{equation}
    \Phi(\alpha_1,\alpha_2,\alpha_3)
= \frac{g'_{\alpha_1}(g_{\alpha_3}-g_{\alpha_2}) - g'_{\alpha_2}(g_{\alpha_3}-g_{\alpha_1})}{\alpha_1-\alpha_2}D
- \frac{32N\sin(\alpha_1-\alpha_2)\cos(\alpha_3)}{\alpha_1-\alpha_2}
\end{equation}
We decompose $\Phi(\alpha_1,\alpha_2,\alpha_3) = \Phi_1(\alpha_1,\alpha_2,\alpha_3) - \Phi_2(\alpha_1,\alpha_2,\alpha_3)$, with
\begin{equation}
    \Phi_1(\alpha_1,\alpha_2,\alpha_3)
= \frac{g'_{\alpha_1}(g_{\alpha_3}-g_{\alpha_2}) - g'_{\alpha_2}(g_{\alpha_3}-g_{\alpha_1})}{\alpha_1-\alpha_2}D,
\quad
\Phi_2(\alpha_1,\alpha_2,\alpha_3)
= \frac{32N\sin(\alpha_1-\alpha_2)\cos(\alpha_3)}{\alpha_1-\alpha_2}
\end{equation}
We observe
\begin{equation}
    \min N = 3 - \max \, g_{\alpha_1}g_{\alpha_2} = 2 > 0,
\qquad
\min D = 16(\sqrt{2}-1) > 0,
\end{equation}
hence $\Phi_2>0$ for $\alpha_1 \neq \alpha_2$.
Now, on rearranging $\Phi_1(\alpha_1,\alpha_2,\alpha_3)$ as,
\begin{equation}
    \Phi_1(\alpha_1,\alpha_2,\alpha_3)
= g_{\alpha_3}\frac{g'_{\alpha_1}-g'_{a_2}}{\alpha_1-\alpha_2}
+ \frac{g_{\alpha_1}g_{\alpha_2}}{\alpha_2-\alpha_1}\left[\frac{g'_{\alpha_2}}{g_{\alpha_2}} - \frac{g'_{\alpha_1}}{g_{\alpha_1}}\right]
\end{equation}
and defining $h_x = \frac{g'_{x}}{g_{x}}$ then
\begin{equation}
    \Phi_1(\alpha_1,\alpha_2,\alpha_3)
= g_{\alpha_3}\frac{g'_{a_1}-g'_{a_2}}{\alpha_1-\alpha_2}
+ g_{\alpha_1}g_{\alpha_2}\frac{h_{\alpha_2}-h_{\alpha_1}}{\alpha_2-\alpha_1}
\end{equation}
By the Mean Value Theorem,
\begin{equation}
    \frac{g'_{\alpha_1}-g'_{\alpha_2}}{\alpha_1-\alpha_2} = g''_{\xi_1},
\quad
\frac{h_{\alpha_2}-h_{\alpha_1}}{\alpha_2-\alpha_1} = h'_{\xi_2},
\end{equation}
for some $\xi_1,\xi_2\in(\alpha_1,\alpha_2)$.
Since
\begin{equation}
    g''_{x} = -(1+\sqrt{2})(\sin x + \cos x) < 0, \qquad
h'_{x}
= \frac{g_{x}g''_{x}-\big(g'_{x}\big)^2}{g_{x}^2} < 0
\end{equation}
we conclude $\Phi_1(\alpha_1,\alpha_2,\alpha_3)<0$.
Thus $\Phi(\alpha_1,\alpha_2,\alpha_3)<0 \implies \mathcal{F}(\alpha_1,\alpha_2,\alpha_3) \neq 0$, for any $\alpha_1 \neq \alpha_2$, and there are no critical points with $\alpha_1 \neq \alpha_2$.
So we can set $\alpha_1 = \alpha_2= \alpha$. Then
\begin{equation}
    s_2(\alpha,\alpha_3) = \frac{N(\alpha,\alpha_3)}{D(\alpha,\alpha_3)}
\end{equation}
where $N(\alpha,\alpha_3) = 3 - g_{\alpha}^2 + 2 g_{\alpha} g_{\alpha_3}$, and $D(\alpha,\alpha_3) = 16\left(\sqrt{2}-\cos(\alpha_3) + \sin 2\alpha \sin(\alpha_3)\right)$. The critical points of $s_2$ satisfy
\begin{equation}
    \nabla s_2 = 0
\quad\Longleftrightarrow\quad
(\nabla N)D - (\nabla D)N = 0.
\end{equation}
The partial derivatives are
\begin{equation}
    \partial_\alpha N = 2 g'_{\alpha}\bigl[g_{\alpha_3}-g_{\alpha}\bigr],
\qquad
\partial_{\alpha_3} N = 2 g_{\alpha} g'_{\alpha_3},
\end{equation}
\begin{equation}
    \partial_\alpha D = 32 \cos(2\alpha)\sin(\alpha_3),
\qquad
\partial_{\alpha_3} D = 16\bigl[\sin(\alpha_3) + \sin(2\alpha)\cos(\alpha_3)\bigr]
\end{equation}
The $\alpha$--component of the critical point condition reads
\begin{equation}
    (\partial_\alpha N)D - (\partial_\alpha D)N = 0,
\end{equation}
which, after substitution and rearrangement, becomes
\begin{equation}
g'_{\alpha}\bigl[g_{\alpha_3}-g_{\alpha}\bigr]\bigl[\sqrt{2}-\cos(\alpha_3) + \sin(2\alpha)\sin(\alpha_3)\bigr]
=
\cos(2\alpha)\sin(\alpha_3)\,
\bigl[3 - g_{\alpha}^2 + 2 g_{\alpha} g_{\alpha_3}\bigr]
\label{eq:xeq}
\end{equation}
On the domain $(0,\tfrac{\pi}{4})$ we have
\begin{equation}
    \cos(2\alpha) > 0,\quad \sin(\alpha_3) > 0,\quad
\sqrt{2} - \cos(\alpha_3) + \sin(2\alpha)\sin(\alpha_3) > 0,
\end{equation}
and the term in bracket on the right-hand side is strictly positive.
Hence the right-hand side of Eq. \eqref{eq:xeq} is positive throughout the domain.
If $\alpha>\alpha_3$, then $g_\alpha>g_{\alpha_3}$ and $g'_\alpha>0$, implying $g'_{\alpha}\bigl[g_{\alpha_3}-g_{\alpha}\bigr] < 0,$
which contradicts Eq. \eqref{eq:xeq}. Therefore, no critical points exist in the
region $\alpha>\alpha_3$.
\subsubsection*{Reduction to $0<\alpha<\alpha_3$}
We now restrict attention to $\alpha_3 \in \left(0,\tfrac{\pi}{4}\right), \alpha \in \left(0,\alpha_3\right)$, and introducing the variables
\begin{equation}
    u_\alpha = \sin \alpha + \cos \alpha,
\qquad
u_{\alpha_3} = \sin\alpha_3 + \cos\alpha_3,
\qquad
v_\alpha = \cos \alpha - \sin \alpha.
\end{equation}
Then $ \sin(2\alpha) = u_\alpha^2 - 1, \cos(2\alpha) = u_\alpha v_\alpha,$ and the function $g_\alpha$ satisfy
\begin{equation}
    g_\alpha = \gamma (u_\alpha-1),
\qquad
g'_\alpha = \gamma v_\alpha,
\qquad
g_{\alpha_3} = \gamma \bigl(u_{\alpha_3}-1\bigr),
\end{equation}
where $\gamma = 1+\sqrt{2}$. Substituting these expressions into Eq. \eqref{eq:xeq} and canceling the positive
factor $v_\alpha$, we obtain
\begin{equation}
\gamma (u_{\alpha_3}-u_{\alpha})\bigl[\sqrt{2}-\cos(\alpha_3) + (u_{\alpha}^2-1)\sin(\alpha_3)\bigr]
=
u_{\alpha}\sin(\alpha_3)
\Bigl[
3 - \gamma^2 (u_{\alpha}-1)^2 + 2\gamma^2 (u_{\alpha}-1)(u_{\alpha_3}-1)
\Bigr]
\label{eq:uxeq}
\end{equation}
The domain becomes $ u_{\alpha_3}\in \left(1,\sqrt{2}\right), u_{\alpha}\in \left(1,u_{\alpha_3}\right)$. Then from Eq. (\ref{eq:mathcalF})
\begin{equation}
    \mathcal{F}(u_{\alpha},\alpha_3)
    =
    \gamma (u_{\alpha_3}-u_{\alpha})\, \mathcal{M}(u_{\alpha})
    - u_{\alpha}\sin(\alpha_3)\, \mathcal{N}(u_{\alpha})
\label{eq:F}
\end{equation}
where
\begin{equation}
    \begin{aligned}
        \mathcal{M}(u_{\alpha}) = \sqrt{2}-\cos(\alpha_3) + (u_{\alpha}^2-1)\sin(\alpha_3), \qquad
\mathcal{N}(u_{\alpha}) = 3 - \gamma^2 (u_{\alpha}-1)^2 + 2\gamma^2 (u_{\alpha}-1)(u_{\alpha_3}-1)
    \end{aligned}   
\end{equation}
Then Eq. \eqref{eq:uxeq} is equivalent to $\mathcal{F}(u_{\alpha},\alpha_3) = 0$.
Note that $\mathcal{M}(u_{\alpha})$ and $\mathcal{N}(u_{\alpha})$ are positive on the domain. Now, as $u_{\alpha} \to u_{\alpha_3}$,
\begin{equation}
    \mathcal{F}(u_{\alpha_3},\alpha_3)
=
- u_{\alpha_3}\sin(\alpha_3)\bigl(3+\gamma^2 (u_{\alpha_3}-1)^2\bigr) < 0.
\end{equation}
Differentiating Eq. (\ref{eq:F}) with respect to $u_\alpha$ gives
\begin{equation}
   \frac{d\mathcal{F}}{du_{\alpha}}
=
\alpha\bigl[-\mathcal{M}+(u_{\alpha_3}-u_{\alpha_1})\mathcal{M}'\bigr]
-
\sin(\alpha_3)\bigl[\mathcal{N}+u_{a_1} \mathcal{N}'\bigr]
\label{eq:dF/dux}
\end{equation}
where
\begin{equation}
    \mathcal{M}'(u_{\alpha}) = 2u_{\alpha}\sin(\alpha_3) > 0,
\qquad
\mathcal{N}'(u_{\alpha}) = -2\gamma^2(u_{\alpha}-u_{\alpha_3}) > 0
\quad\text{for } u_{\alpha} < u_{\alpha_3}.
\end{equation}
on substituting $\mathcal{M}'$ and $\mathcal{N}'$ is Eq. (\ref{eq:dF/dux}), we get
\begin{equation}
    \frac{d\mathcal{F}}{du_{\alpha}} = -\left[
\mathcal{M}\gamma + \mathcal{N}\sin(\alpha_3) + 2\gamma u_{\alpha}(u_{\alpha_3}-u_{\alpha})(\gamma-1)\sin(\alpha_3)
\right]
\end{equation}
It follows that $\frac{d\mathcal{F}}{du_{\alpha}}<0$ for all $u_{\alpha} \in (1, u_{\alpha_3})$, and hence $\mathcal{F}$ is
strictly decreasing in $u_{\alpha}$. Since $\mathcal{F}(u_{\alpha_3},\alpha_3) < 0$, even as $u_{\alpha} \rightarrow u_{\alpha_3}$ we conclude that
\begin{equation}
    \mathcal{F}(u_{\alpha},\alpha_3) < 0
\quad\text{for all } u_{\alpha}\in(1,u_{\alpha_3}),
\end{equation}
and therefore the equation $\mathcal{F}(u_{\alpha},\alpha_3) = 0$ has no interior solution.
Combining with the case $\alpha > \alpha_3$, we conclude that $s_2$ has no interior critical
points.
\subsection*{Boundary analysis and conclusion}
Since $s_2(\alpha_1,\alpha_2,\alpha_3)$ admits no interior critical points, all extrema must occur on the
boundary of the domain, this leaves us with the corner points
\begin{equation}
    s_2\left(\frac{\pi}{4},\frac{\pi}{4},\frac{\pi}{4}\right) = \frac{1}{4\sqrt{2}}
\quad\text{and}\quad
s_2\left(0,0,0\right)=\frac{3(1+\sqrt{2})}{16}
\end{equation}
Imposing the condition $\lambda_{01}(\alpha_1,\alpha_2,\alpha_3) > 0$ for all $\alpha_1,\alpha_2,\alpha_3$ yields the
admissible range
\begin{equation}
    s \in \left(0,\frac{1}{4(\sqrt{2}+1)}\right]
\;\cup\;
\left[\frac{3(1+\sqrt{2})}{16},\infty\right).
\end{equation}
The interval $\left(0,\frac{1}{4(\sqrt{2}+1)}\right]$ is excluded since the self-testing statement in Eq.~(\ref{eq:f_beta}) implies that to obtain non trivial fidelity the observed violation $\beta^* = 4\sqrt{2}-\frac{1}{2s} = 2(\sqrt{2}-1)$, which is less than the classical bound of $4$ for the $3-$ partite Svetlichny operator Hence the smallest admissible value is
\begin{equation}
    s_* = \frac{3(1+\sqrt{2})}{16}.
\end{equation}
A similar analysis for the remaining $\lambda_{x_1x_2}(\alpha_1,\alpha_2,\alpha_3)$ shows
that they are also strictly positive for this choice of $s_*$ and the corresponding $\mu = -\frac{2+3\sqrt{2}}{4}$. This completes
the proof.
}

\end{widetext}
\end{document}